\documentclass{aa}    % LaTeX A&A  Standard Fonts

\def\simless{\mathbin{\lower 3pt\hbox
     {$\rlap{\raise 5pt\hbox{$\char'074$}}\mathchar"7218$}}} %< or of order
\def\simgreat{\mathbin{\lower 3pt\hbox
     {$\rlap{\raise 5pt\hbox{$\char'076$}}\mathchar"7218$}}} %> or of order

\newcommand{\Lsun} {L$_\odot$}
\newcommand{\Msun} {M$_\odot$}

\usepackage {psfig, epsfig}

\begin{document}

%\thesaurus{02.16.2, 13.09.1, 11.09.4, 11.13.2, 11.09.1}

\title{ISOCAM survey and dust models of 3CR radio galaxies and
  quasars\thanks{Based on observations with ISO, an ESA project with
instruments funded by ESA Member States (especially the PI countries:
France, Germany, the Netherlands and the United Kingdom) with the
participation of ISAS and NASA.}}

\author {R.~Siebenmorgen\inst{1}
        \and W.~Freudling\inst{2,1}
        \and E.~Kr\"ugel\inst{3}
        \and M.~Haas\inst{4}}

\institute{European Southern Observatory, Karl-Schwarzschildstr. 2, 
        D-85748 Garching b. M\"unchen
\and
        Space Telescope -- European Coordinating Facility,
        Karl-Schwarzschildstr. 2, D-85748 Garching b. M\"unchen
\and 
        Max-Planck-Institut for Radioastronomy, Auf dem H\"ugel 69,
        Postfach 2024, D-53010 Bonn
\and 
        Astronomisches Institut, Ruhr-Universit\"at, 
        Universit\"atsstr. 150, D-44780   Bochum}

\offprints{rsiebenm@eso.org} \date{Received 22 December 2003 / 
Accepted 18 March 2004}

\abstract{We present a survey of all 3CR sources imaged with ISOCAM
onboard the {\it Infrared Space Observatory (ISO)}.  The sample
consists mostly of radio--loud active galactic nuclei (AGN).  For each
source, we present spatially integrated mid--infrared (MIR, $5 -
18\mu$m) fluxes measured from newly calibrated ISOCAM images.  In
total, we detected 68 objects of the 3CR catalogue, at redshifts $z
\le 2.5$, and obtained upper limits for 17 objects.  In addition, we
detected 10 galaxies not listed in the 3CR catalogue.  The one with
the highest redshift is 4C$+$72.26 at $z = 3.53$.  ISOCAM data are
combined with other photometric measurements to construct the spectral
energy distribution (SED) from optical to radio wavelengths.  The MIR
emission may include synchrotron radiation of the AGN, stars of the
host galaxy or dust.  Extrapolation of radio core fluxes to the MIR
show that the synchrotron contribution is in most cases negligible.
In order to describe dust emission we apply new radiative transfer
models.  In the models the dust is heated by a central source which
emits photons up to energies of 1keV. By varying three parameters,
luminosity, effective size and extinction, we obtain a fit to the SED
for our objects.  Our models contain also dust at large (several kpc)
distance from the AGN.  Such a cold dust component was neglected in
previous computations which therefore underestimated the AGN
contribution to the far infrared (FIR).  In 53 cases ($\sim 75$\,\% of
our detected 3CR sources), the MIR emission can be attributed to dust.
The {\it hot dust} component is mainly due to small grains and PAHs.
The modelling demonstrates that an AGN heating suffices to explain the
ISO broad band data, starburst activity is not necessary.  In the
models, a type 1 AGN is represented by a compact dust distribution,
the dust is therefore very warm and emission of PAHs is weak because
of photo--destruction.  In AGNs of type 2, the dust is relatively
colder but PAH bands are strong.
	}

\maketitle

\section{Introduction}

More than 20 years ago Kotanyi \& Ekers (1979) pointed out that for
radio--loud elliptical galaxies, the orientation of the (optical) dust
lane appears preferentially perpendicular to the radio axis.  In
unification theories dust plays an important role.  In all of them, a
supermassive blackhole accretes gas from a disk and, in radio--loud
objects, a jet is ejected parallel to the rotation axis.  The
blackhole and the disk are surrounded by a torus of interstellar gas
and dust which may block light from the central region on the way to
the observer.  The diversity in the appearance of activity types is
then explained foremost as a result of different viewing angles
(Barthel 1989).

%At least, the schism between Seyfert galaxies of type 1 and 2 is
%convincingly resolved this way.  For instance, the most famous
%example, NGC1068, is classified as Sy~II, but shows in
%spectropolarimetry broad lines characteristic of Sy~I.  This becomes
%understandable if the engine is surrounded by a thick torus viewed
%from the side: the lines from the obscured broad line region are
%scattered by electrons above and below the torus and become therefore
%polarized (Antonucci \& Miller 1985).  Moreover, Fe K$\alpha$ line
%emission at about 6.5\,KeV from highly X--ray ionized iron is detected
%at a level much above what one expects from the observed X--ray flux
%(Krolik \& Kallmann 1987; Koyama et al.~1989).  Again, the discrepancy
%is resolved if one assumes that iron of cosmic abundance (and all in
%the gas phase) is ionized by radiation escaping from the central
%engine in a cone parallel to the torus axis.  Should the hard X--ray
%K$\alpha$ photons cross part of the torus on their way to the
%observer, the X--ray flux will be weakened.

The obscuring dust torus is at the heart of unification theories.  It
reaches towards the blackhole as close as dust can survive ($T
\simless 1500$\,K).  As the torus is not directly seen on optical
images, its dimensions are at most a few hundred parsec. Because of
the immense bolometric luminosity emitted from the AGN, the dust in
the torus, even at a distance of 100\,pc, must be very warm ($ >
100$\,K), and further in even {\it hot}.  Such hot dust will emit
preferentially in the mid infrared (MIR), at wavelengths covered by
the ISOCAM filters.

Dust in radio galaxies is not restricted to the nuclear region, but is
ubiquitous in the host galaxy.  It absorbs and scatters the light, but
also radiates thermally as it is heated either by the nucleus or by
stars.  Of course, the diffuse and spatially extended dust is much
colder than the dust near the AGN.  Rapid star formation is also a
potential heating source, besides the AGN. A detailed search for dust
in the nuclear regions of 120 radio sources of the 3CR catalogue was
carried out by de Koff et al.~(2000) using optical HST images at
0.7$\mu$m center wavelength and of $0.1{''}$ resolution.  In one out
of three galaxies, they found evidence for dust obscuration, with a
large variety of morphologies such as disks, lanes or filaments.  Dust
distributed smoothly on a large scale of several kpc is difficult to
detect in absorption maps but reveals itself by infrared emission.

The dust shroud of an AGN is often not transparent and to model the
emission spectrum one has to compute the radiative transfer.  Models
for optically thick dusty nuclei and galaxies have been carried out in
various approximations (Pier \& Krolik 1993, Laor \& Draine 1993
Kr\"ugel \& Siebenmorgen 1994, Granato \& Danese 1994, Efstathiou
\& Rowan--Robinson 1995, Siebenmorgen et al. 1997, Nenkova et
al. 2002, Popescu et al. 2004). Previous AGN dust models
underestimated the FIR as they did not include dust located at large
distances from the center and thus missed the cold component.  To
overcome the deficit emission in the FIR, these authors add an
additional component to the models and attribute it to starburst
activity (e.g.~Rowan--Robinson 2000).  In Section~4, we present AGN
dust models where large scale dust emission is incorporated and where
the geometry is radically simplified so that they can be described by
only three parameters.

A cold dust component was discovered in 18 sources of the 3CR
catalogue already by IRAS (Heckman et al. 1994) and more recently, in
more objects, by ISOPHOT (Fanti et al.~2000, Meisenheimer et al.~2001,
Van Bemmel et al.~2001, Andreani et al. 2002, Spinoglio et al.~2002,
Haas et al. 2004).  These authors detected also a few sources in the
MIR.  The hot dust responsible for it must be close to the nucleus and
its detection serves as a further test for the unified model
hypothesis.  This motivated us to survey the 3CR galaxies in the MIR
by means of ISOCAM images available from the ISO archive.  The
sensitivity of ISOCAM is $2 - 3$ orders of magnitude greater than IRAS
at 12$\mu$m.

In Section~2 and 3, we present the images, pertaining photometric
measurements as well as results.  The spectral energy distributions
(SED) are displayed in Section~4.  To interprete them, we apply
radiative transfer models.  A discussion of generic SED properties in
the infrared for different AGN types is given in Section~5.  The
conclusions are given in Section~6.  In Appendix~{\ref{galnotes.ap} we
remark on individual galaxies (previous indications of dust, X-ray
properties, jets, presence of high excitation lines, source
morphology, companions).  In Appendix~\ref{fits.ap}, we make for each
galaxy a note on the radiative transfer models of Section~4.  Details
of the ISOCAM observational set up is summarised in
Tab.~\ref{tab.obs}.  The fully reduced and astrometrically corrected
ISOCAM images are shown as overlays on optical images in
Fig.~\ref{images}.

\section {Survey data}

We present the data of our survey, describe which ISOCAM fields were
extracted from the ISO archive, briefly overview the ISOCAM observing
modes, and outline data reduction, source identification and
photometric procedures.

\subsection{Source selection}
\label{ident.sec}

The parent catalogue for our sample is the third update of the revised
third Cambridge catalogue (3CR) by Spinrad et al.~(1985).  It contains
298 sources of which 195 are classified as radio galaxies and 53 as
radio quasars.  Most of the objects in the 3CR catalogue are well
observed at optical and near infrared wavelengths. We searched in the
ISO archive for ISOCAM (Cesarsky et al. 1996) observations made with
the long wavelength array (LW).  All ISOCAM images containing,
potentially, at least, one 3CR source have been extracted. Altogether,
we found 146 such ISOCAM observations. Often the same source was
observed in multiple filter band passes. On the fully reduced ISOCAM
images, we determined and improved the astrometric pointing by
identifying objects with accurate coordinates listed in the SIMBAD
Astronomical Database or the NASA/IPAC Extragalactic Database (NED).
The images are presented in Fig.~\ref{images}. We also made 10
serendipitous detections of galaxies which are not members of the 3CR
catalogue. They are eye-ball detections at the position of NED
coordinates.

\subsection{Observations}

Most of the observations were done in the so called mini-raster mode
where different ISOCAM detector pixels saw the same part of the sky
(see ISO Handbook for a description of observing modes, Leech \&
Pollock, 2001). In addition, there are nine staring and one beam
switch observation.  Three observations were performed in the circular
variable filter (CVF) scan mode and one in the polarimetric mode.

Depending on the lens used in the experiment, the pixel scale of the
32 $\times$ 32 element detector was 1.5$''$, 3$''$ and 6$''$ and the
total field of view $48 \times 48$, $96 \times 96$ and $180 \times
180$\,arcsec$^2$, respectively. The most frequently applied broad band
filter was LW10 with a bandwidth from 8--15$\mu$m and central
wavelength at 12$\mu$m. In addition there are observations with
filter: LW1 (4--5, 4.5$\mu$m), LW2 (5--8.5, 6.7$\mu$m), LW3 (12--18,
14.3$\mu$m), LW7 (8.5--11, 9.6$\mu$m) and CVF scans (5--17$\mu$m) at
spectral resolution of $\lambda/\Delta \lambda \approx 50$. In a
single observing template often more than one filter was selected.

\begin{table}
\caption{ISOCAM photometry of 3CR sources}
\label{tab.flux}
\begin{tabular}{lrrrr}
\hline
\hline
& & & & \\
Name  & TDT & Band & Flux & RMS \\
 &  &$\mu$m  & mJy  & mJy \\
  & & & & \\
\hline
1 &2 &3 &4 &5  \\
\hline
  & & & & \\
3C006.1  & 70101081 & 12.0 & 0.9  & 0.2  \\
3C013    & 80801283 & 12.0 & 0.7  & 0.2 \\
3C017    & 57502102 & 12.0 & 4.0  & 0.4   \\
3C018    & 61901003 & 12.0 & 9.1  & 0.6 \\
3C020    & 59702305 &  12.0 & 5.3 & 0.4   \\
3C022    & 78500882 &  12.0 & 7.1 & 0.4   \\
3C031    & 58703801 &   4.5 &30 & 2   \\
         & 40201422 &   6.7 &36 & 2   \\
         & 58703793 &  12.0 &25 & 1  \\ 
         & 58703801 &  14.3 &30 & 2   \\
 3C033.1 & 59702607 &  12.0 &18 & 1   \\
3C048    & 43901804 &  14.3 &59 & 4   \\
3C061.1  & 56201411 &  12.0 & 4.3 & 0.3 \\
 3C66B   & 45304902 &4.5 &14 & 1   \\
         & 43502724 &6.7 &13 & 1   \\
         & 45304902 &  14.3 & 3.8 & 1.6\\   
3C071 	 & 63301602 & 4.5 &  6860 & 340 \\
(NGC1068)& 63301902 & 6.8 &  14030 & 700   \\
         & 63302202 & 14.9 & 50830 & 2540   \\
 3C076.1 & 46601603 &4.5 & 5.1 & 1.0   \\
         & 46601603 &6.7 & 3.0 & 0.8   \\
         & 46601603 &  14.3 & 1.0 & 1.5  \\ 
3C079    & 61503807 &  12.0 &21 & 2 \\
 3C083.1 & 65901304 &4.5 &31 & 2   \\
         & 65901032 &6.7 &47 & 3   \\
         & 65901304 &  14.3 & 9.7 & 1.5\\   
3C084    & 61503617 &6.7 &  230 &20 \\
         & & 9.6  &  400   &   70 \\
         & &14.3  &  1000   & 100 \\
3C098    & 63302405 &4.5 & 6.5 & 1.0  \\ 
         & 63302405 &6.7 & 7.7 & 0.9   \\
         & 63302405 &  14.3 &24 & 2   \\
3C231    & 12300106  & 6.7  & 25000 & 1250 \\
(M82)    & 12300106  & 12.0 & 63000 & 3150   \\
         & 12300106  & 14.9 & 68000 & 3400   \\
3C249.1  & 21305001 & 12.0  & 19 & 1 \\
3C265    & 22400201 &4.5 & 1.3 & 0.5   \\
         & 22201802 &6.7 & 2.4 & 0.2   \\
         & 22400303 &  12.0 & 6.0 & 1.6  \\ 
3C270    & 22801205 &4.5 &  160 & 8  \\
         & 22801205 &6.7 &95 & 5   \\
         & 22801205 &  14.3 &38 & 3  \\ 
 3C272.1 & 23502406 &4.5 &  900 &  46 \\
         & 23100414 &6.7 &  240 & 12   \\
         & 23100414 &9.6 &  150 & 8   \\
         & 23100414 &  14.3 &53 & 3   \\
3C273    & 24100504 & 14.3 & 290 & 15 \\
	 & 24100504 &  6.7 & 190 & 10 \\
3C274    & 23800308 &4.5 &  740 & 38 \\
(M87)    & 23800308 &6.7 &  260 & 13   \\
         & 23901834 &  12.0 &51& 3   \\
3C277.3  & 24500106 &  6.7 & 3.6 & 0.3  \\
         & 24500106 &  14.3 & 8.5 & 0.5   \\
 &  & & & \\
\hline
\end{tabular}
\end{table}

\setcounter{table}{0}
\begin{table}
\caption{ - continued.}
\begin{tabular}{lrrrr}
\hline
\hline
& & & & \\
Name  & TDT & Band & Flux & RMS \\
 &  &$\mu$m  & mJy  & mJy \\
  & & & & \\
\hline
1 &2 &3 &4 &5  \\
\hline
  & & & & \\
3C286	& 38800808 & 12.0 & 3.6& 0.3  \\
3C288.1 & 24407233 &  12.0 & 2.3 & 0.6   \\
3C293   & 61701307 &4.5 &16 & 1   \\
        & 61701307 &6.7 &14 &  1   \\
        & 61701413 &  12.0 &19 & 2  \\ 
        & 61701307 &  14.3 &23 & 1   \\
3C295   & 18001405 &6.7 & 1.4 & 0.3 \\
3C296   & 27000606 &4.5 &30 & 2    \\
        & 27000606 &6.7 &16 & 1   \\
        & 27000606 &  14.3 &13 & 1  \\ 
3C303.1 & 52900115 &  12.0 & 1.9 & 0.2 \\
3C305   & 46300408 &4.5 &12 & 1   \\
        & 12300205 &6.7 & 7.1 & 2.7   \\
        & 51400760 &  12.0 &21 & 1   \\
        & 12300205 &  14.3 &27 & 5   \\
3C305.1 & 71702771 &  12.0 & 1.5 & 0.2   \\
3C309.1 & 60001472 &  12.0 & 8.2 & 0.5   \\
        & 60092200 &  6.7 & 4.4 & 0.5   \\ 
3C319   & 54100619 &  12.0 & 1.3 & 0.2   \\
3C321	& 65800208 &  6.7 & 12 &  1    \\
	& 65800208 & 12.0 & 27 &  1    \\
	& 65800208 & 14.4 & 51 &  3    \\
3C324   & 30300612 &6.7 & 0.1 & 0.2 \\
        & 30300413 &  12.0 & 1.7 & 0.2   \\
3C330   & 33600323 &6.7 & 1.3 & 0.2   \\
        & 60001373 &  12.0 & 1.7 & 0.2   \\
        & 33600323 &  14.3 & 2.6 & 0.5   \\
3C332   & 60201725 &  12.0 & 9.2 & 0.5 \\
3C336   & 30400442 &  12.0 & 2.4 & 0.5   \\ 
        & 30492200 &  6.7 & 0.9 & 0.2   \\ 
 3C338  & 10601408 &4.5 &16 & 1  \\
        & 10601408 &6.7 & 9.4 & 0.8   \\
        & 51100561 &  12.0 & 8.8 & 0.5   \\ 
        & 10601408 &  14.3 & 7.3 & 1.4   \\
3C341   & 60201626 &  12.0 & 8.0 & 0.5   \\
 3C343  & 61900565 &  12.0 & 1.4 & 0.2   \\
  3C345 & 51100679 &  12.0 &22 & 1   \\
3C346   & 64001327 &  12.0 & 4.2 & 0.3  \\ 
3C349   & 52500429 &  12.0 & 3.4 & 0.3   \\
3C351   & 52500130 &  12.0 &32 & 2   \\
3C356   & 52100668 &  12.0 &0.6 & 0.2  \\ 
3C371   & 52901232 &  12.0 &66 & 4   \\
        & 55500476 &  14.3 & 91 & 5   \\
3C379.1 & 52901433 &  12.0 & 3.6 & 0.3   \\ 
3C380   & 54100969 &  12.0 &14 & 1   \\
3C381   & 50304334 &  12.0 &19 & 1   \\
3C382   & 52500562 &  12.0 &85 & 5   \\
3C386   & 47100710 &4.5 &13 & 1   \\
        & 47100710 &6.7 & 6.3 & 0.8   \\
        & 51301035 &  12.0 & 5.6 & 0.3  \\ 
        & 47100710 &  14.3 & 3.3 & 1.4   \\
3C388   & 54100836 &  12.0 & 1.2 & 0.2 \\
3C390.3 & 52901537 &  12.0 &94 & 5    \\
 & &  & & \\
\hline
\end{tabular}
\end{table}

\setcounter{table}{0} 
\begin{table}
\caption{ - continued.}
\begin{tabular}{lrrrr}
\hline
\hline
& & & & \\
Name  & TDT & Band & Flux & RMS \\
 &  &$\mu$m  & mJy  & mJy \\
  & & & & \\
\hline
1 &2 &3 &4 &5  \\
\hline
  & & & & \\
3C402   & 49604639 &  12.0 & 3.4 & 0.5   \\
3C410   & 53503041 &  12.0 &22 & 1   \\
3C418   & 54101870 &  12.0 & 3.6 & 0.3 \\  
3C430   & 54301344 &  12.0 & 1.7 & 0.3   \\
3C433 & 52201845 &  12.0 & 32.0 & 2 \\
3C445 & 54500518 &6.7 &93 &5\\
      & 54500518 &  14.3 &  210 & 11   \\
3C449 & 54801112 &4.5 &16 & 1 \\
      & 36701710 & 6.7 &10 & 1   \\
      & 37403041 &  12.0 & 7.6 & 0.9 \\
      & 54801252 &  12.0 & 8.7 & 0.5 \\  
      & 36701710 &  14.3 &11 & 2   \\
3C452 & 54801053 &  12.0 &23 & 1 \\
3C456 & 54700854 &  12.0 & 9.1 & 0.6    \\
3C459 & 37500303 & 4.5  & 0.1 & 0.5 \\
      &          & 6.0  & 5.5 & 0.9 \\
      &          & 7.7  & 6.8 & 1.1 \\
      &          & 14.9  & 29.5 & 4.0 \\
3C465 & 39501112 &4.5 &22 & 1   \\
      & 39501112 &6.7 & 8.1 & 0.8   \\
3C469.1 &56201064 &  12.0 & 1.8 & 0.2 \\
 & & & & \\
\hline
\end{tabular}
\end{table}

\begin{table} 
\caption{ISOCAM $3\sigma$ upper limits of 3CR sources}
\label{tab.upper}
\begin{tabular}{lrrr}
\hline
\hline
  &  &   & \\
Name  & TDT & Band & Flux \\
      &     & $\mu$m& mJy \\
  &  &  &   \\
\hline 
1 &2 &3 &4  \\
\hline 
  &  &  &   \\
3C002   & 41907203 &  12.0 &  $<$1.7 \\
3C313   & 08800387 &  12.0 &  $<$6.2 \\
3C314.1 & 52900417 &  12.0 &  $<$0.6 \\
3C320   & 59501620 &  12.0 &  $<$0.5 \\
3C343.1 & 39901248 &  12.0 &  $<$3.3 \\
3C352   & 51800667 &  12.0 &  $<$0.6 \\
3C357   & 51800831 &  12.0 &  $<$0.6 \\
3C368   & 71200283 &  6.7  &  $<$1.5  \\
        & 10200421 &  6.7  &  $<$8.1 \\
        & 71200283 &  14.3 &  $<$2.1 \\
3C401   & 53200838 &  12.0 &  $<$0.6 \\
3C427.1 & 52902663 &  12.0 &  $<$0.6 \\
3C434   & 53503146 &  12.0 &  $<$1.0 \\
3C436   & 52201948 &  12.0 &  $<$1.0 \\
3C437   & 53503288 &  12.0 &  $<$1.4 \\
3C438   & 54000949 &  12.0 &  $<$0.5 \\
3C442   & 53702911 &  14.3 &  $<$4.5 \\  
3C454.1 & 56001074 &  12.0 &  $<$0.6 \\
3C460   & 56501357 &  12.0 &  $<$0.4 \\
  &  &  &  \\
\hline
\end{tabular}
\end{table}

Parameters of the observing templates are detailed in
Tab.~\ref{tab.obs}: target dedicated time (TDT) by which the
observations are identified (column~1), celestial coordinate
of the map center (column~2 and 3), observing date (column~4), band
pass filter (column~5), lens (column~6), number of raster points $M$
(column~7) and raster lines $N$ (column~8), step between raster points
$dM$ (column~9) and raster lines $dN$ (column~10), both in arcsec, and
the median number of exposures taken on each raster position
(column~11) together with the integration time per read--out
(column~12).

\subsection{Data reduction}

The data were reduced with the ISOCAM Interactive Analysis (CIA, Ott
et al.~1996). We used the default data reduction steps of CIA: dark
current subtraction, initial removal of cosmic ray hits (glitches),
detector transient fitting, exposure coaddition, flat fielding and
flux conversion to astronomical units. Except for staring modes, the
coadded images at each raster position were sky projected and
corrected for field distortion.

The dark current depends on the orbital position of the ISO
space-craft and the temperature of the ISOCAM detector.  The applied
correction is based on a model described by Roman \& Ott (1998). The
deglitching is done by following the temporal signal variation of a
pixel using a multi--resolution wavelet transform algorithm (Starck et
al.~1997).  The response of the detector pixels depends on the
previous observations and there may be long term hysteresis effects
for each detector element after changes of the photon flux level.  We
applied the detector flux transient fitting method for ISOCAM data
(Coulais \& Abergel 2000).  After application of the default
deglitcher some residuals of cosmic ray impacts were still visible in
the data.  Therefore we applied after the detector flux transient
correction a second cosmic ray rejection method which is basically a
multi-sigma clipping of the temporal signal (Ott et al.~2000).  For
raster observations, flat field estimation can be improved by an
iterative method exploiting the fact that the same sky area is seen by
different detector pixels (Starck et al.~1999).  The method works best
for highly redundant data where a large number of pixels see the same
sky.  In case of low redundancy rasters and for staring observations,
we flat--fielded according to the standard calibration library.

\subsection{Source photometry} 
\label{photo.sec} 

To determine the source flux, we perform a multi--aperture photometry.
For each aperture centered on the brightest pixel of the source we
determine the background as the mean flux derived in a 4 pixel wide
annulus which is put 2 pixels away from the greatest ($\sim 18''$)
aperture. In this way, we obtain aperture fluxes that flatten with
increasing aperture radius and approach an asymptotic value. For
point sources the same procedure is repeated on a theoretical and
normalised point source image (Okumura 1998) to find the correction
factor of the multi--aperture analysis. The correction factor is
typically 5\%. The determination if a source is a point source or not
is done by eye. The procedure will introduce an error of less than the
correction factor for slightly extended sources misjudged to be a
point source. In order to derive the statistical flux uncertainty, for
the source aperture we quadratically add the RMS image and the
1$\sigma$ uncertainty of the background estimate.  The systematic
error of ISOCAM photometry is typically $\simless 5\%$ and color
corrections of order 10\% (see ISOCAM Handbook, Blommaert et
al.~2001).

\section{Results}

We find from the detection statistics of the ISOCAM survey of 3CR
sources that in our sample there is no bias against AGN type.  We
present SEDs from optical to radio wavelengths for individual sources
and discuss potential contributions to the ISOCAM fluxes (synchrotron,
photospheric and dust emission).

\begin{table} 
\caption{ISOCAM photometry of galaxies not in the 3CR catalogue}
\label{tab.fluxother}
\begin{tabular}{llrrr}
\hline
\hline
& & & & \\
Name  & TDT & Band & Flux & RMS \\
 &  &$\mu$m  & mJy  & mJy \\
  & & & & \\
\hline
1 &2 &3 &4 &5  \\
\hline
  & & & & \\
4C+72.26 & 74101506 &  12.0 & 0.7 & 0.08   \\
 NGC0382 & 58703801 &4.5 & 7.0 & 1.2   \\
         & 40201422 &6.7 &12 & 2   \\
         & 58703793 &  12.0 & 7.3 & 0.5   \\
         & 58703801 &  14.3 &10 & 2   \\
 NGC7236 & 53702911 &6.7 & 3.8 & 0.8 \\
 &53702750 &  12.0 & 2.8 & 0.3   \\
 &53702911 &  14.3 & 5.5 & 1.5   \\
 NGC7237 & 53702911 &6.7 & 3.6 & 0.8 \\
 &53702750 &  12.0 & 1.8 & 0.2   \\
 NGC5532NED02 & 27000606 &  4.5  &4.5 & 1.0 \\
       &27000606 &  6.7  &1.0 & 0.8   \\
 F17130+5021 & 52100804 & 12.0 & 14 & 1 \\
J1728201+314603 & 51800831 & 12.0 & 2.8  & 0.2	     \\
PMNJ0214$-$11582 &  59601514 &  12.0 & 1.8 & 0.3 \\
 PGC012293 & 65901032 &6.7 & 1.3 & 0.2   \\
    &65901304 &  14.3 & 3.0 & 1.4   \\
  PCd91212130.83$+$ & 35701635 &6.7 &31 & 2   \\
\quad \quad 245139.2 & & & \\ 
\hline
\end{tabular}
\end{table}

\subsection{Bias and detection statistics}

The primary goal of our survey is to compile a sample of radio-loud
AGNs which is not subjects to biases related to the type and therefore
viewing angle of the AGN torus. While this seems to be the case for
the parent 3CR catalog (e.g. Urry and Pdovani 1995), the selection of
objects from the ISOCAM survey was beyond our control. We shall
therefore compare the flux and redshift distribution of the galaxies
detected by ISOCAM with those of the 3CR catalogue.

\begin{figure}[htb]
\centerline{\epsfig{file=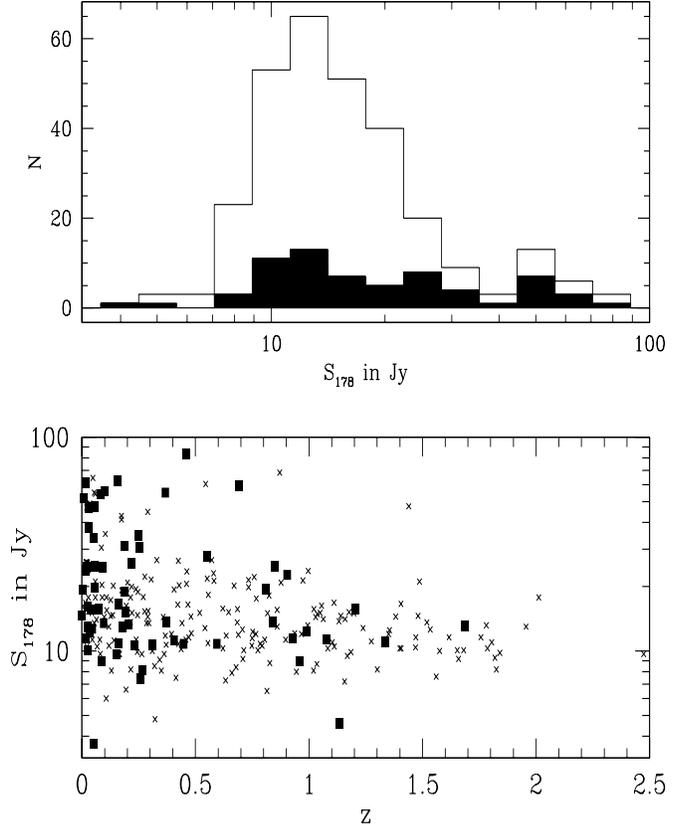,width=9.cm,height=12cm} }
\caption{Top: Histogram of the distribution of the 178\,MHz radio
fluxes in the 3CR catalogue.  The filled part represents the objects
detected by ISOCAM.  Bottom: Radio fluxes of 3CR sources at 178\,MHz
plotted as a function of redshift.  The filled squares are objects of
this paper.  }
\label {3c.select}
\end{figure}

\begin{figure}[htb]
\centerline{\epsfig{file=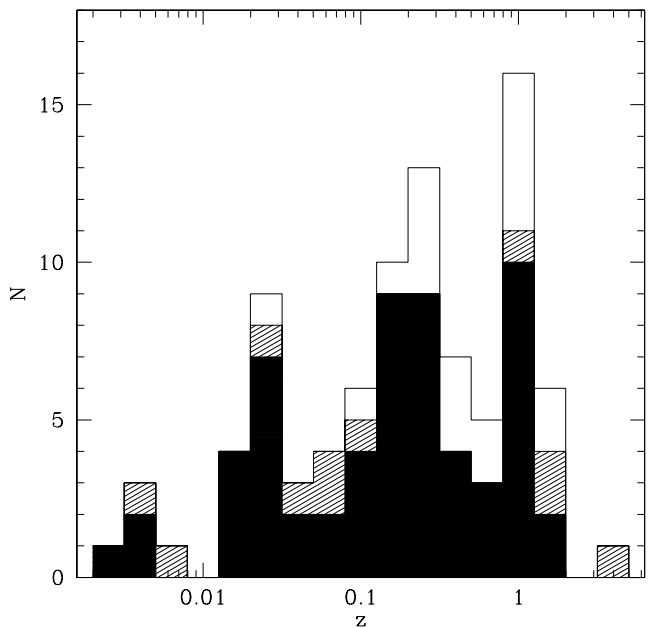,width=9.cm} }
\caption{Detection rate of 3CR sources in our survey as a function of
redshift.  The filled part part of the columns refers to 3CR sources
detected with ISOCAM where the MIR emission can be attributed to dust
(see Sect.\ref{sec.modelfit}).  The shaded part represents all ISOCAM
detections of the 3CR catalogue and the upper (blank) histogram shows
the redshift distribution of 3CR sources sampled by ISOCAM. }
\label {3c.complete}
\end{figure}

In total, we detected 68 sources from the 3CR catalogue and determined
17 upper limits up to redshifts $z \leq 2.5$.  Furthermore, there
are 10 serendipitous detections of galaxies in the ISOCAM images
which are not listed in the 3CR catalogue.  One example is the radio
source 4C$+$72.26 at redshift $z = 3.53$.  The ISOCAM fluxes of
detected 3CR sources are listed in Tab.~\ref{tab.flux} and upper
limits of non-detected 3CR sources in Tab.~\ref{tab.upper}.  We also
give the ISOCAM flux for galaxies which are not members of the 3CR
catalogue in Tab.~\ref{tab.fluxother}.  Notations in
Tab.~\ref{tab.flux} and Tab.~\ref{tab.fluxother} are: catalogue name
(column~1), the TDT number used as an identifier of the observation is
listed in column~2, followed by the central wavelength ($\mu$m) of the
filter band pass (column~3), aperture flux (column~4) and 1$\sigma$
RMS flux error (mJy) of the aperture photometry (column~5).
Tab.~\ref{tab.upper} has the same structure except that we specify as
flux the 3$\sigma$ upper limit.  Notes to 3CR galaxies detected by
ISOCAM are given in Appendix~\ref{galnotes.ap}.

In Fig.~\ref{3c.select} we compare the radio flux and redshift
distribution of 3CR sources detected with ISOCAM to galaxies of the
parent catalogue. From the top panel of Fig.~\ref{3c.select} it
becomes evident that the fraction of 3CR galaxies detected in
our survey is a strong function of radio flux, which spans two orders
of magnitude.  For example of the sources with an 178\,MHz flux
$\simless 10$\,Jy we detect only 15\%, whereas for sources with an
178\,MHz flux stronger than 30\,Jy $\sim 50\%$ of all 3CR sources are
detected.  Furthermore, the galaxies detected with ISOCAM follow, by
and large, the redshift--flux distribution of the parent catalogue
(lower box of Fig.~\ref{3c.select}), at least, for redshifts larger
than about 0.2.  However, at $z < 0.2$, our subsample seems to show a
preponderance of radio bright objects.

The fact that low redshift objects are preferentially included raises
the question whether this is due to flux limits of our survey,
i.e.~whether the ISOCAM detection rate strongly drops as a function of
redshift.  This is investigated in Fig.~\ref{3c.complete}, where we
present the detection rate of our survey as a function of redshift.
From Fig.~\ref{3c.complete} one can see that the detection rate of
local objects ($z < 0.1$) is close to 100\%. For $z> 0.2$, and then
almost independent of redshift up to $z \sim 1$, the detection rate is
$\sim$75\%. Therefore our survey is in this redshift range not limited
by the sensitivity of ISOCAM. We conclude that the AGNs of our survey
present a fair subsample of the 3CR catalogue for redshifts larger
than 0.2.

\subsection{Spectral energy distributions}
\label{sec.comparison}

For each detected 3CR source we compiled photometric measurements from
optical to radio wavelengths.  Our primary source of information was
the NASA/IPAC Extragalactic Database (NED).  In addition, we availed
ourselves of the IRAS fluxes (re-processing raw data with XSCANPI, see
NASA/IPAC home page), the ISOPHOT fluxes (Fanti et al.~2000,
Meisenheimer et al.~2001, Van Bemmel et al.~2001, Andreani et al. 2002,
Spinoglio et al.~2002, Haas et al.~2004), mm/submm continuum data
(Kr\"ugel et al.~1990, Gear et al.~1994, Steppe et al.~1988, Steppe et
al.~1995, Best et al.~1998, Cimatti et al.~1998, Robson et al.~2001,
Haas et al. 2004) and 1.4\,GHz fluxes from the NRAO VLA Sky Survey
(Condon et al.~1998).  The resulting observed SEDs are shown in
Fig.~\ref{sed.fig}, together with additional information discussed
below.

The MIR emission of our radio galaxies may come from synchrotron
radiation of the AGNs, stars within the galaxies and dust emission.
We assume that bremsstrahlung can be neglected.  To identify a
synchrotron component at ISOCAM wavelengths, we extrapolated radio
core fluxes of the 3CR sources with a power law or, if that was not
possible, we assumed a spectral slope as derived from global radio
properties.  The latter include the core emission (a few arcsecs in
size) and the radio lobes which may extend several arcmin.  The ISOCAM
observations have a resolution of $\sim 10''$ and are centered on the
AGNs.  Consequently, one has to use radio core fluxes (Giovannini et
al.~1988; Hardcastle \& Worrall~1999) to avoid an overestimate of the
synchrotron component.  When such core fluxes were not available, we
adopted the core--to--total radio power correlation by Giovannini et
al.~(2001) to derive some estimate.  As can be seen from
Fig.~\ref{sed.fig}, the synchrotron contribution to the MIR emission
is generally small.  One exception is the BL Lac 3C371, which will be
neglected in the following.

We assess the significance of the stars by fitting a redshifted $T =
4000$\,K blackbody to the optical and near infrared data and
extrapolate it to the MIR.  The effective temperature was chosen
deliberately low to maximise the effect of the stars at MIR
wavelengths.  The further interpretation of this curve depends on the
size of the optical relative to the ISOCAM image.  When they are
comparable, we directly subtracted from the ISOCAM flux the
extrapolated stellar flux.  When the optical observation comprised a
much larger region, the stellar component was considered as the upper
limit of a stellar component within the ISOCAM aperture. The
stellar component is simply added to the emission of the galaxy
without any radiative transfer calculations. Indeed, for some of our
sources the ISOCAM flux is dominated by stars from the host galaxy
(see, for instance, the SED of 3C270 in Fig.~\ref{sed.fig}). To
quantify the likely contribution of the dust emission to the ISOCAM
fluxes, we present below radiative transfer models of dust enshrouded
AGNs.

\section {Radiative transfer in dusty AGN}
\label{sec.modelfit}

\begin{figure*}
\epsfig{file=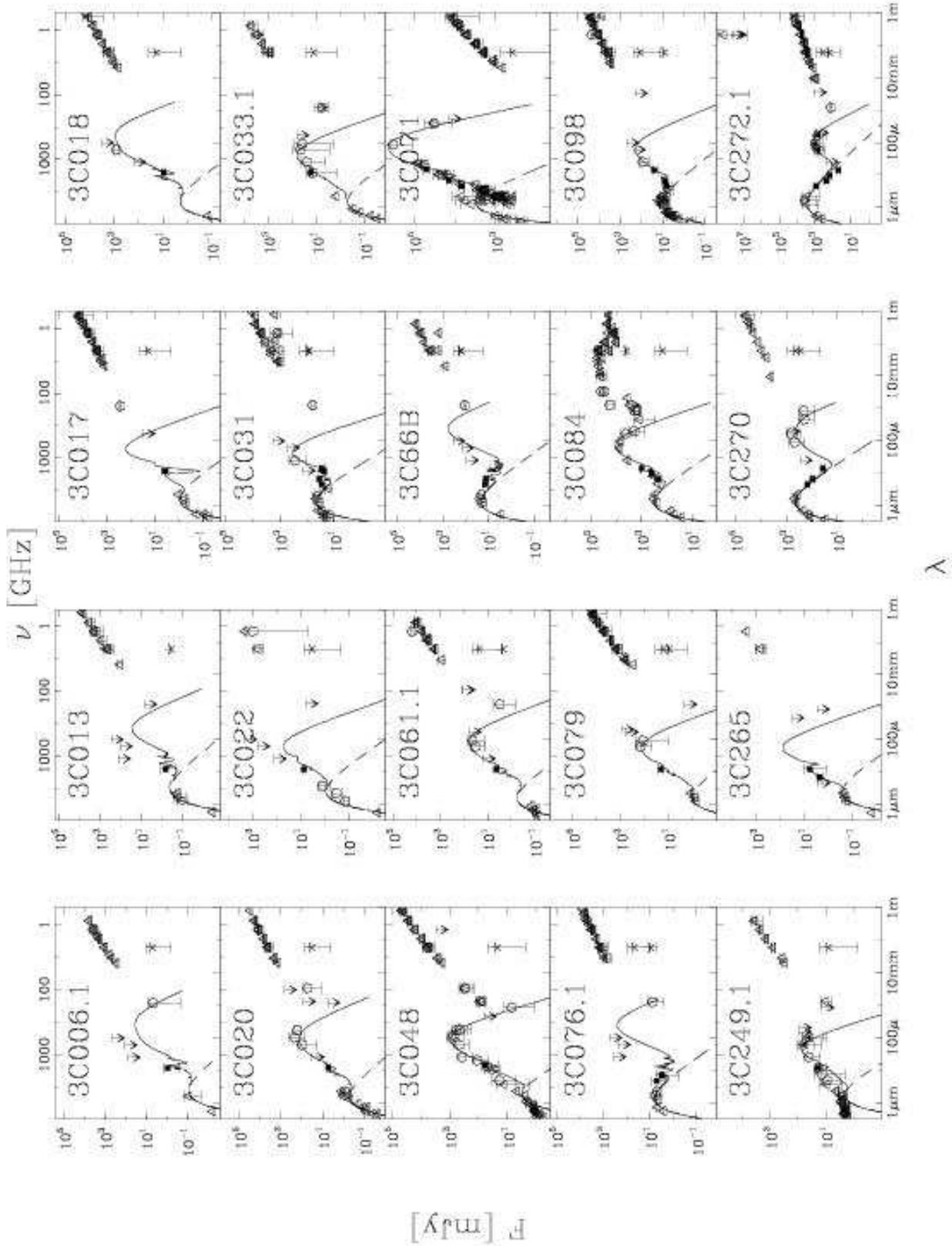,width=17cm}
\caption{Observed spectral energy distributions of the sources of the
ISOCAM survey: ISOCAM photometry (filled squares), NED data
(triangles), submillimeter data (pentagons), radio core fluxes
(asterisks).  Error bars are 1$\sigma$ and upper limits 3$\sigma$
RMS. Solid lines represent the addition of the photospheric component
(dashed lines) and the dust emission resulting from the AGN models.}
\label{sed.fig}
\end{figure*}

\setcounter{figure}{2}
\begin{figure*}
\epsfig{file=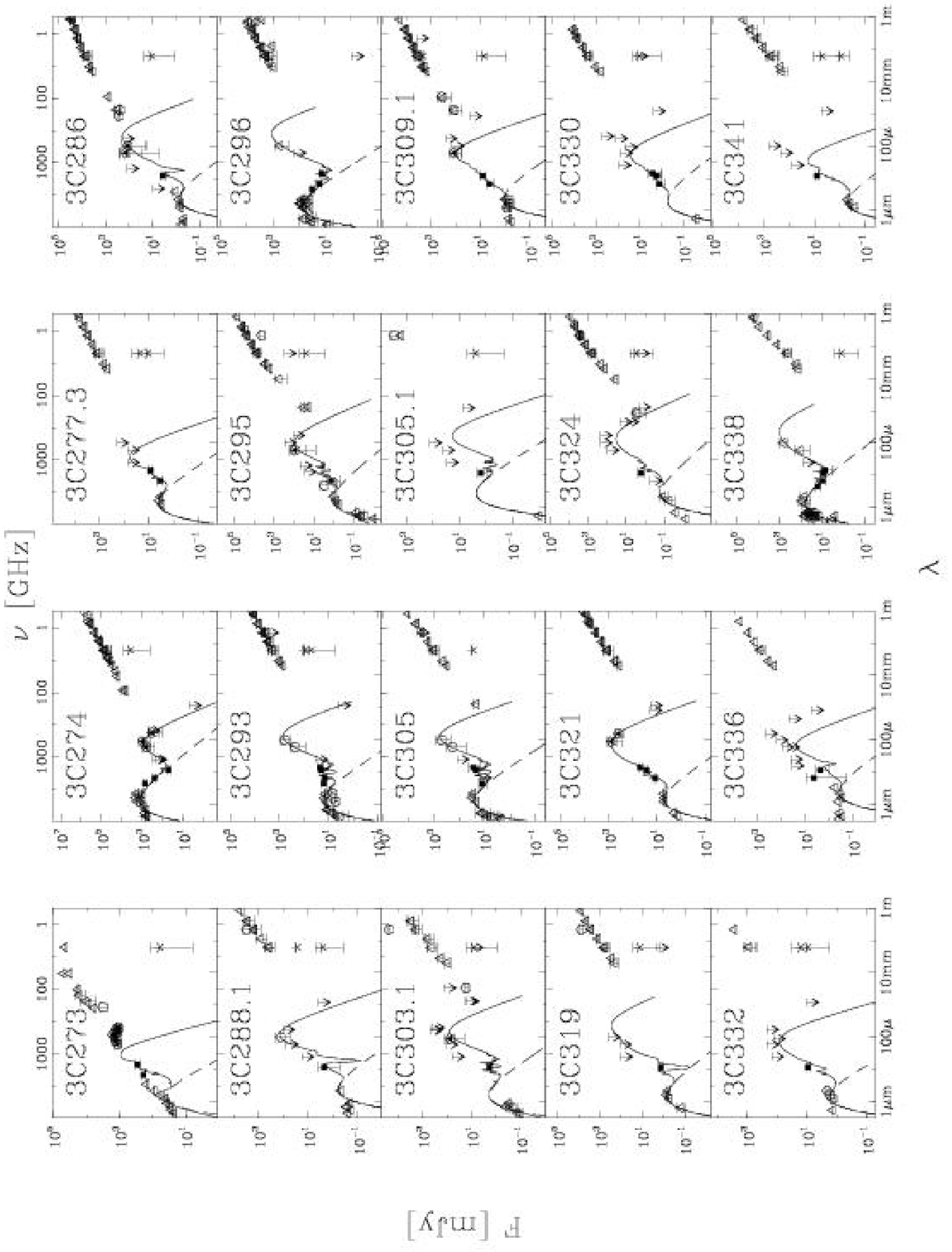,width=17cm}
\caption{ -- continued --}
\end{figure*}

\setcounter{figure}{2}
\begin{figure*}
\epsfig{file=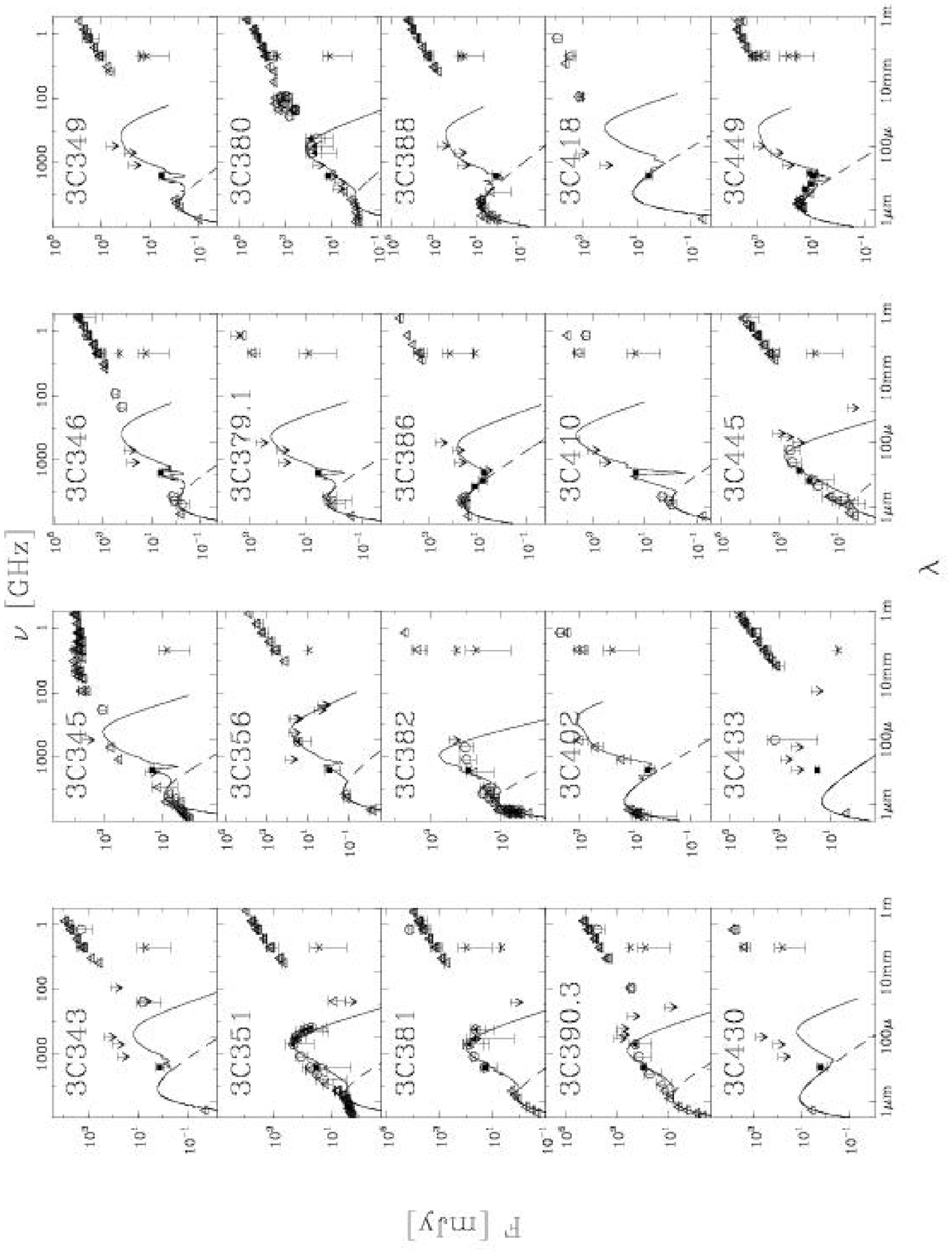,width=17cm}
\caption{ -- continued --}
\end{figure*}

\setcounter{figure}{2}
\begin{figure*}
\centerline{\hspace{-0.5cm}
\psfig{file=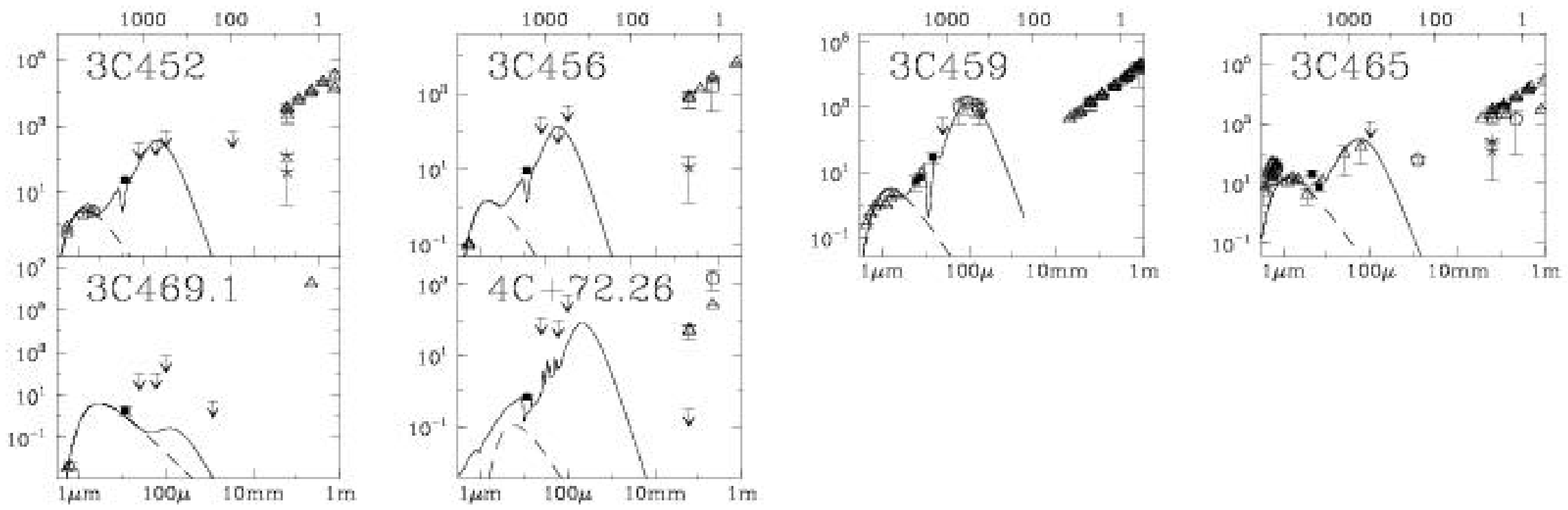,width=15.2cm,angle=-90}}
\vspace{-8.5cm}
\caption{ -- continued --}
\end{figure*}

Here we outline an AGN dust model which is based on three free
parameters.  This simple model seems sufficient to fit the dust
emission of the observed SED.  After some motivation we present the
model of the central engine and of the dust, the parameters and a
library of reference SEDs. Then we discuss model properties and the
fit to the data.  The computation of the dust emission together with
the procedure to consider photospheric and synchrotron components
provides a tool to specify a likelihood of the detection of dust in
radio -- loud AGN. Notes on fits of individual sources are given in
Appendix~\ref{fits.ap}.

\subsection{Motivation}

In most 3CR objects, the mid and far IR flux cannot arise from stars
nor from the radio core because an extrapolation of either component
to the infrared fails by orders of magnitude.  This confirms the
widely held view that the MIR/FIR emission of 3CR sources is often
dominated by the dust (Meisenheimer et al.~2001, Van Bemmel et
al.~2001, Andreani et al. 2002, Freudling et al. 2003, Haas et
al. 2004). As the sources are not optically thin, one has to somehow
compute the radiative transfer.  In our procedure we
self--consistently solve the radiative transfer in spherical symmetry
with a ray--tracing method.  The heating source of the dust is the
central engine, scattering by dust is taken into account.  Absorption
and scattering efficiencies as well as asymmetry factors of the grains
are calculated from Mie theory.  The size distribution which the dust
grains display is approximated by 8 different grain sizes for carbon
and silicate material, respectively.  Collisional heating (Contini \&
Contini 2003) is disregarded.  The dust model includes a component of
tiny and transiently heated particles.  Further details can be found
in Siebenmorgen et al. (1992).  Observations pertaining to the
distribution of dust in 3CR galaxies were performed by Martel et
al.~(1999) using the HST.  Most galaxies could be resolved and dust
was found close to the central heating source as well as far (few kpc)
from it.  Their images imply a complex morphology of the dust
obscuration which defies any simple mathematical description.  To be
computational tractable, a radical simplification of the geometry is
needed. As there are only few observational infrared data points for
each galaxy, a sophisticated source structure that would require for
its specification more parameters than available data points is not
warranted.  We therefore consider spherical symmetry.

This is, of course, not in agreement with the unified scheme where the
AGN is surrounded by an accreting disk or a torus and the
phenomenological type is determined by the viewing angle of the
observer.  Such a configuration is intrinsically, at least,
axisymmetric.  However, the strongest effect brought about by the
torus on the emerging spectrum occurs, of course, at visual and
ultraviolet wavelengths.  In the infrared, which we try to reproduce,
the distance of the dust to the source and its mass are more
important.  Therefore, when barring the short wavelengths, a sphere
can still conserve key elements of a multi--dimensional object.  We
note also that the angular (not the radial) distribution of the dust
is in radiative transfer calculations only of importance when the
source is opaque.

The dust distribution in the model is a spherical average over the
true distribution.  When applied to fitting an observed SED, such a
model will tend to overestimate the near infrared emission of a type 2
AGN (edge--on viewed disk), and underestimate it for a type 1
(pole--on oriented disk).  Nevertheless, as viewing angles cover a
continuous range, in a statistical analysis, one should get about the
right amount of cold and hot dust as well as an estimate of their
distances from the central heating source.

Another reason to apply here spherically symmetric models stems from
the well known difficulty which disks have in explaining the
10$\mu$m--silicate band.  This resonance lies in a spectral region
covered by ISOCAM.  When viewed face--on, disks should show the
feature in emission, contrary to what is observed.  Spherical models,
on the other hand, are not heavily plagued by this problem as there is
always some extinction towards the center.  As a third argument we
point out that for one of the best known AGNs, in the Circinus galaxy,
a spherical model proves as efficient in reproducing the observed
infrared SED and the wavelength dependence of the infrared sizes
(Siebenmorgen et al.~1997) as do two--dimensional models (Ruiz et
al.~2001).  The reason lies in the fact that most infrared data are of
low spatial resolution.

Previous AGN torus models (e.g. Pier \& Krolik 1993, Efstathiou \&
Rowan--Robinson 1995) do not include the dust in the host galaxy.
Such models therefore miss the cold dust altogether.  As the apertures
of FIR observations are large one has to artifically include in
these models a cold component (Farrah et al. 2003), whereas we
incorporate it energetically self--consistent.  Finally, the dust
distribution as derived from HST images (de Koff et al. 2000) has so
far never revealed a perfect torus like structure.

\subsection{The central engine}

The central engine is approximated by a (small) sphere with a power
law monochromatic luminosity $L_\nu \propto \nu^{-0.7}$ in the range
from 10\AA \/ to 2$\mu$m.  The exact lower limit, $\lambda_{\rm low}$,
should not be important as the dust becomes transparent (Dwek \& Smith
1996).  The upper limit, $\lambda_{\rm up}$, is equally uncertain and
represents a synchrotron break.  Fortunately, the choice of
$\lambda_{\rm low}$ and $\lambda_{\rm up}$ has little influence on the
final IR spectrum.

\subsection{The dust model}

The dust consists of large carbon and silicate grains with radii, $a$,
between 300 and 2400\AA \/ and a size distribution $n(a) \propto
a^{-3.5}$; very small graphites ($a = 10$\AA); and two kinds of PAHs
(30 C and 20 H atoms; 252 C and 48 H atoms) which are responsible for
various infrared emission bands (see Siebenmorgen et al.~(2001) and
Siebenmorgen et al.~(2004) for details of the dust model).

The contribution of the different dust populations to the AGN spectrum
is studied in Fig.~\ref{dustcomponents.fig}.  For fixed intrinsic
luminosity of the AGN $L = 10^{12}$ \Lsun, radius $R =1$\,kpc and
dust extinction $A_{\rm {V}} = 8$\,mag we show the SED for three dust
models: i) large grains only; ii) large grains and small graphites;
iii) all dust components are included (large grains, small graphites
and PAHs).  Of particular interest for the interpretation of our
ISOCAM observations is the MIR.  Here dust emission below 6$\mu$m is
due to small graphites which become very hot ($\simgreat 1000$\,K)
after absorption of a single UV photon.  The emission between 6 and
9$\mu$m and near 11.3 and 12.8$\mu$m is dominated by the PAH bands.
Large grains with temperatures $\sim 300$\,K contribute to the
emission at wavelengths $\sim 10\mu$m.  At FIR/submm wavelengths, the
emission is entirely due to cold large grains ($\simless 50$\,K). In
summary the entire MIR is dominated by the emission of {\it hot dust}
which may be either large grains heated to high temperatures, small
graphites or PAH.

\subsection{Model library}
\label{sec.modpara}

In the spectra displayed in Fig.~\ref{sed.fig}, one can usually make
out three components: optical and near infrared emission, probably of
stellar origin ($\simless 5\mu$m), the mid and far infrared ($5\mu$m
$\simless \lambda \simless 1000\mu$m), and the radio regime ($>
1000\mu$m).  Our fits concern only the second component which we
assume to be due to dust and to originate in the central, few kpc
region of galactic nucleus.  The optical and radio emission come from
totally separated areas and may thus simply be added to the mid and
far IR spectrum.

To keep the models and their subsequent interpretation as simple as
possible, we consider only three basic parameters: the luminosity, the
effective linear size of the galaxy, and the dust (or gas) mass.
Furthermore, we adopt a constant dust density.  We may then
alternatively use as parameters: luminosity $L$, radius $R$, and
visual extinction $A_{\rm V}$.

We build a set of reference models{\footnote {The full set of models
is available at: \newline {\tt
http://eso.org/\~ \,rsiebenm/agn\_models}}} by varying $L$, $A_{\rm V}$
and $R$ in discrete steps over a wide range.  In particular, we change
the luminosity from 10$^{8.5}$ to 10$^{15.25}$ \Lsun, the visual
extinction from 1 to 128\,mag and the outer radius of the interstellar
cloud that surrounds the AGN from 0.125 to 16\,kpc; the inner radius
of the dust shell is chosen to be close to the evaporation radius of
the grains; a typically number is 1\,pc.

\begin{figure}
\centerline{\psfig{file=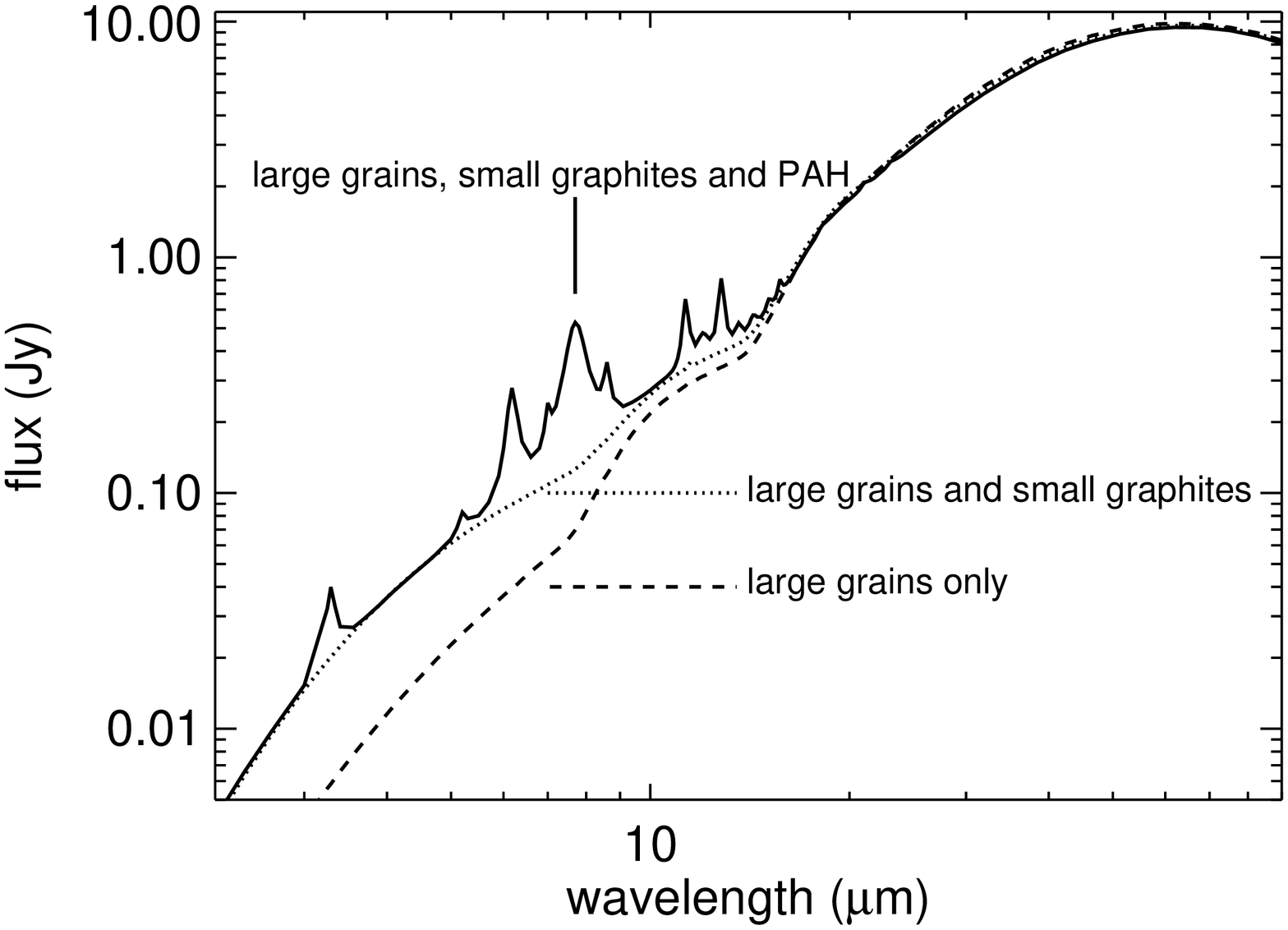,width=7.0cm,angle=90}}
\caption{The influence of the dust population on the infrared spectrum
of the AGN.  Model parameters are: visual extinction $A_{\rm {V}} =
8$\,mag, radius $R = 1$\,kpc,  luminosity $L =
10^{12.5}$ \Lsun \/ at distance of 50\,Mpc.  Model with large grains
only (dashed), with large grains and small graphites (dotted), full
model with large grains, small graphites and PAH (solid line).}
\label{dustcomponents.fig}
\end{figure}

\begin{figure}
\centerline{\psfig{file=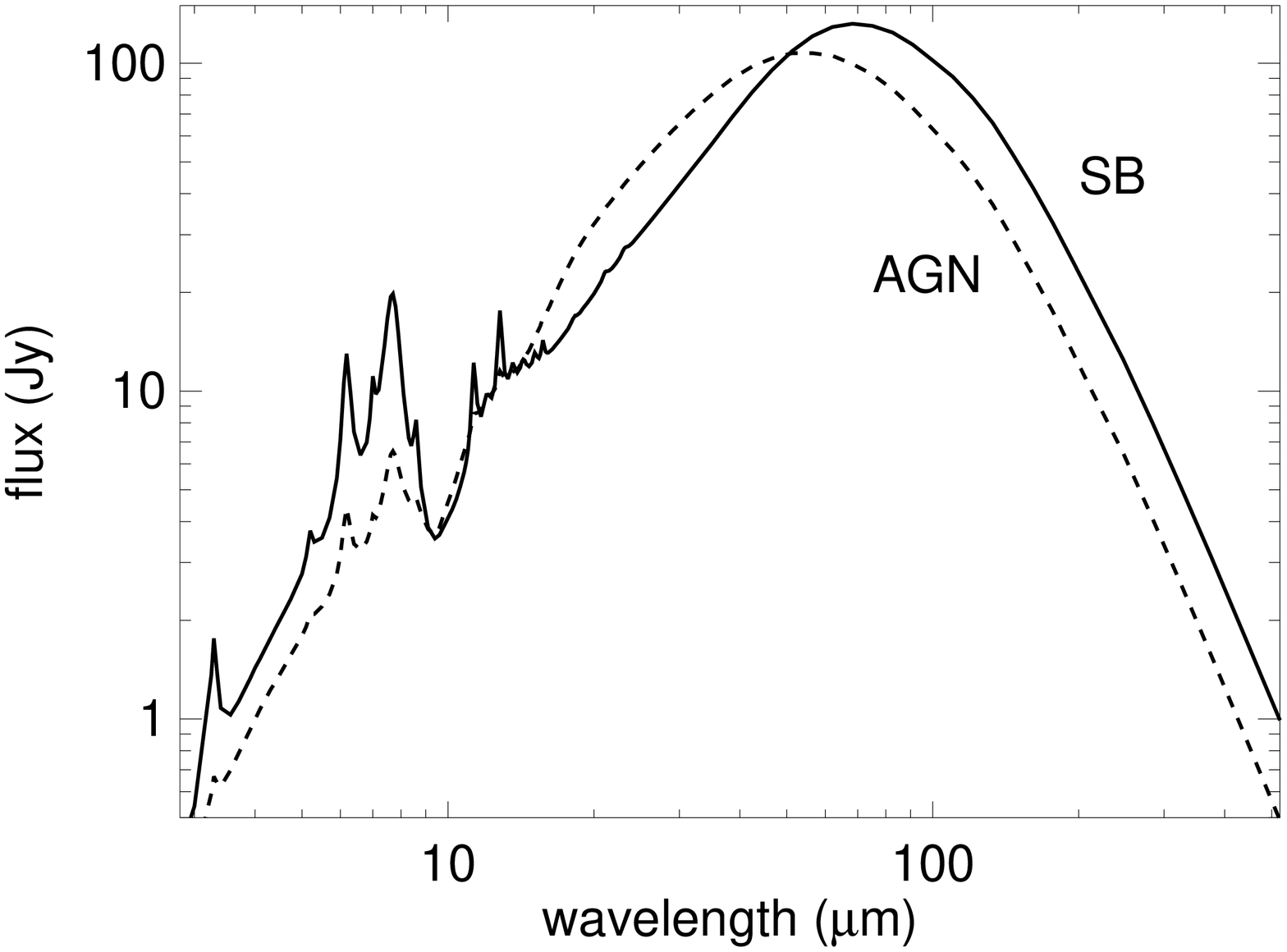,width=7.0cm,angle=90}}
\caption{Comparison of radiative transfer computations of dusty
starbursts (solid) and AGNs (dashed).  In both models we applied a
visual extinction of $A_{\rm {V}} = 32$\,mag, luminosity of $L =
10^{12}$ \Lsun \/ at distance of 50\,Mpc, and radius of $R = 1$\,kpc.}
\label{sb_agn.fig}
\end{figure}

\subsection{Model properties}

In our models, the central engine of the AGN is the sole source for
heating the dust.  Alternatively, the dust could, at least partially,
be heated by a starburst activity.  The influence on the spectrum is
investigated in Fig.~\ref{sb_agn.fig} where we compare an AGN with a
starburst model.  In either model, the visual extinction $A_{\rm {V}}
= 32$\,mag, the luminosity $L = 10^{12}$ \Lsun \/ and the cloud radius
is $R = 1$\,kpc.  In the starburst, all stars have a uniform
luminosity and temperature ($L_* = 20000$\Lsun, $T_{\rm {eff}} =
20000$\,K) and are surrounded by a local dust shell of constant
density 10$^4$\,H-atoms/cm$^3$ (for details of the radiative transfer
in a starburst, see Kr\"ugel \& Siebenmorgen 1994). We see in
Fig.~\ref{sb_agn.fig} that in the SED of a starburst the PAH features
and the submillimeter emission are considerably stronger.  As our
simple AGN model already fit observed spectra of radio--loud quasars
(Freudling et al. 2003) we simplify our procedure and ignore in the
following any starburst component.

Once we have chosen to use AGN models, there are three basic
parameters in our modelling. The AGN luminosity of the models is
in most cases well constrained by the data and corresponds to the
integrated flux. The total extinction determines absorption features
within the spectrum whereas the outer radius of the dust can be used
to balance the relative amount of hot and cold dust for a given
extinction.

The influence of the extinction, $A_{\rm V}$, on the SED is shown in
Fig.~\ref{agn_lav.ps} for three different luminosities and for a fixed
radius of 1\,kpc.  For the same luminosity and radius a high
extinction implies strong submillimeter emission, a deep silicate
absorption feature at 10$\mu$m and highly reddened radiation at
shorter wavelengths.  When the extinction is low, on the other hand,
the AGN becomes optically visible, the silicate feature disappears and
the submillimeter flux is weak.  The SED peak may be used as a measure
of the average dust temperature of the bulk material.  For the same
$R$ and $L$ value a lower $A_{\rm V}$ implies a SED peak at shorter
wavelengths as compared to models with high $A_{\rm V}$.  On average,
the dust is then much warmer. The bulk material becomes, of course,
also warmer when the AGN luminosity increases: models with same
extinction and radius show a SED peak shifted to shorter wavelengths
at higher luminosities (Fig.~\ref{agn_lav.ps}).

\begin{figure} 
\centerline{\psfig{file=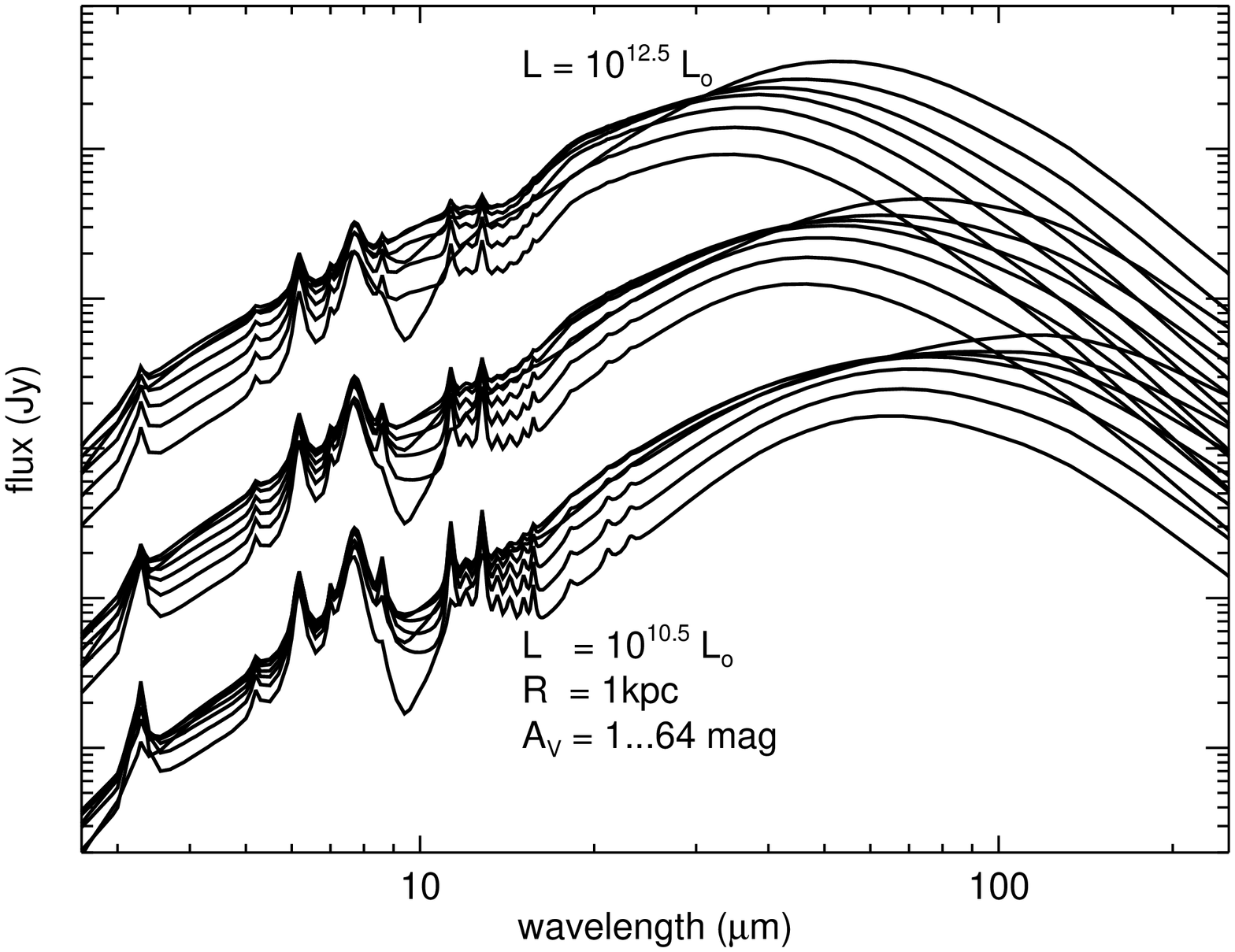,width=7.cm,angle=90}} 
\caption{AGN model spectra are shown for visual extinctions of $A_{\rm
{V}} =$ 1, 2, 4, 8, 16, 32 and 64\,mag and luminosities of $L =
10^{10.5}, 10^{11.5}$ and $10^{12.5}$ \Lsun \/ at a distance of
50\,Mpc. The radius is fixed at $R = 1$\,kpc. }
\label {agn_lav.ps}
\end{figure}

\begin{figure}
\centerline{\psfig{file=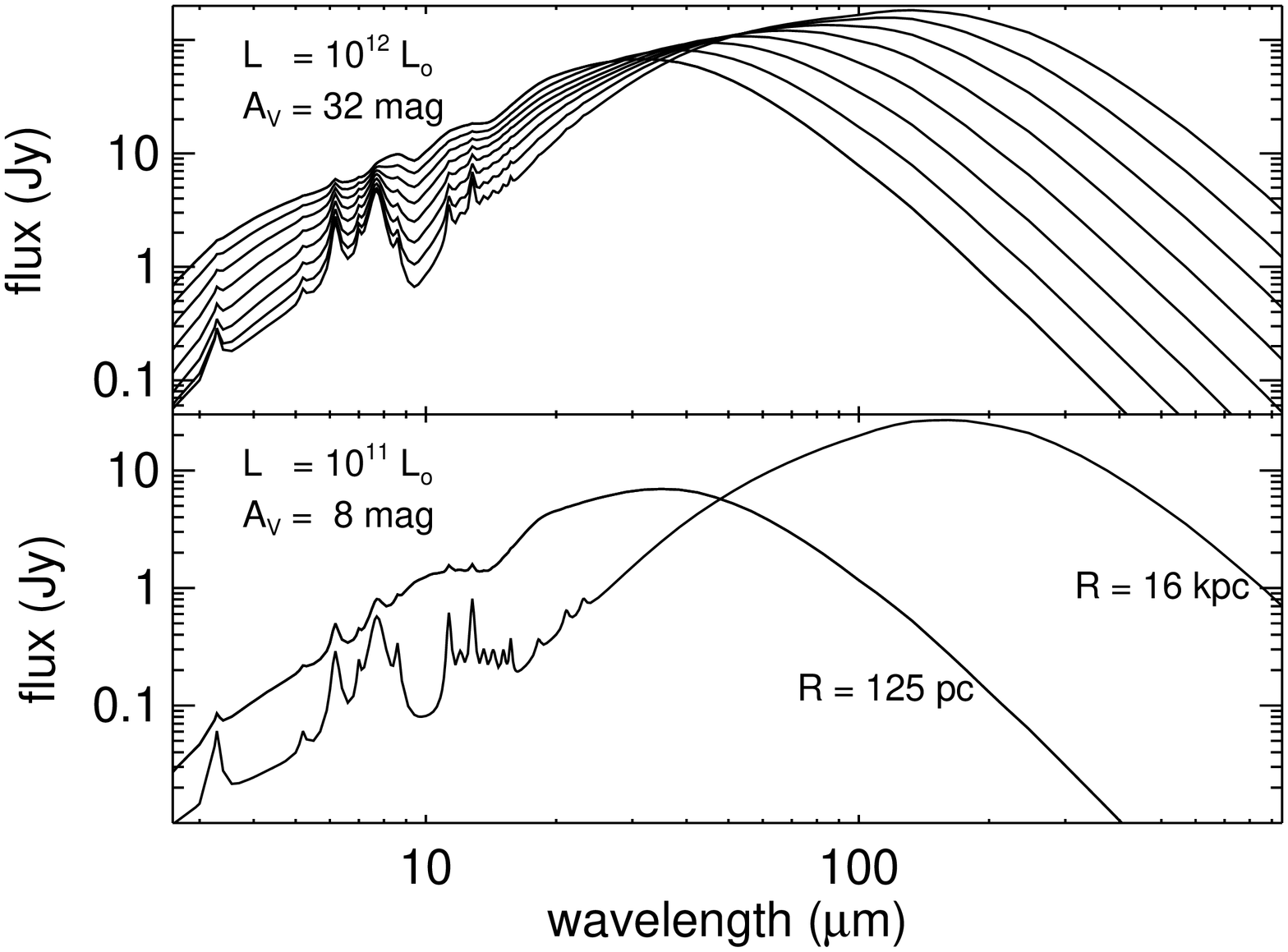,width=9.cm,height=9.5cm,angle=90}} 
\caption{The influence of the radius on the AGN model spectra.  We
apply a distance of 50\,Mpc. Top: For luminosity of $L = 10^{12}$
\Lsun \/ and extinction of $A_{\rm {V}} = 32$\,mag we present models
with $R =$ 0.125, 0.25, 0.5, 1, 2, 4, 8 and 16\,kpc.  Bottom: For
luminosity of $L = 10^{11}$ \Lsun \/ and extinction of $A_{\rm {V}} =
8$\,mag we show models with $R =$ 0.125 and 16\,kpc. }
\label {agn_lrad.ps} 
\end{figure}

\begin{table*}
\caption{Infrared dust luminosity of 3CR radio galaxies. Galaxies
where spectral shape of the dust emission is constrained are marked in
bold.}
\label{tab.dust}
\begin{center}
\begin{tabular}{l c l r r r l}
\hline
\hline
&     & & & & & \\
Name & z  & Type   & log(L$_{\rm dust}$)& $R$ & $A_{\rm V}$ & Dust ? \\
&     & &  [\Lsun] &[pc] &[mag] & \\
&     & & & & & \\
\hline
{\bf 3C006.1} & 0.840 &  NLRG/FRII & 11.75   & 16000 & 32 & certain \\
   3C013 & 1.351 &  NLRG/FRII & 12.25   & 16000 &  2 & likely \\
   3C017 & 0.220 &     BLRG   & 11.5    &   125 &128 & likely \\
   3C018 & 0.188 &     BLRG   & 12      & 16000 & 32 & certain \\
{\bf 3C020} & 0.174 &  NLRG/FRII & 11.25   &  2000 & 16 & certain \\
   3C022 & 0.937 &  BLRG/FRII & 12.5    &   500 &  8 & possible \\
{\bf 3C031} & 0.017 &   LERG/FRI & 10      &   125 &  2 & certain \\
{\bf 3C033.1} & 0.181 &  BLRG/FRII & 11.25   &   125 & 16 & certain \\
{\bf 3C048} & 0.367 &    QSO     & 12.75   &  4000 & 32 & certain \\
{\bf 3C061.1} & 0.186 &  NLRG/FRII & 11   &  1000 & 16 & certain \\
   3C66B & 0.212 &   LERG/FRI &  9.75   & 16000 &  2 & certain \\
{\bf 3C071} & 0.004 &     NLRG   & 11.5    &  2000 & 16 & certain \\
 3C076.1 & 0.032 &   LERG/FRI & $<$9.75 & 16000 &  2 & no evidence \\
{\bf 3C079} & 0.256 &  NLRG/FRII & 11.75   &   500 & 16 & certain \\
{\bf 3C084} & 0.018 &   NLRG/FRI & 11.25   &  2000 &  4 & certain \\
{\bf 3C098} & 0.031 &  NLRG/FRII & 10      &   125 & 32 & certain \\
   3C231 & 0.007 &   LERG/FRI & 10.5    &  1000 & 16 & certain \\
{\bf 3C249.1} & 0.312 &      QSO   & 11.75   &   125 & 32 & certain \\
   3C265 & 0.811 &  NLRG/FRII & 12.5    &   125 & 64 & certain \\
   3C270 & 0.007 &   NLRG/FRI & $<$ 8.75&  2000 & 32 & no evidence \\
 3C272.1 & 0.003 &   LERG/FRI &  9      &   500 &  2 & certain \\
   3C273 & 0.158 &   QSO      & 12.75   &   500 &  1 & possible \\
   3C274 & 0.001 &   NLRG/FRI & $<$ 9   &   250 &  2 & no evidence \\
{\bf 3C277.3} & 0.086 &     BLRG   & 10.5    &   250 &  1 & certain \\
   3C286 & 0.849 &    QSO/    & 12.5    &  8000 & 64 & certain \\
 3C288.1 & 0.961 &      QSO   & 12.5    &  1000 &128 & likely \\
{\bf 3C293} & 0.045 &   LERG/FRI & 10.75   &  8000 &  2 & certain \\
   3C295 & 0.461 &  NLRG/FRII & 12.25   &  8000 &  1 & certain \\
   3C296 & 0.461 &   LERG/FRI & 10      & 16000 &  8 & likely \\
 3C303.1 & 0.267 & NLRG/FRII  & 11.5    & 16000 &  1 & certain \\
{\bf 3C305} & 0.041 &   NLRG/FRI & 10.75   &  8000 &  1 & certain \\
 3C305.1 & 1.132 &     NLRG   & $<$ 12  &  8000 &  2 & no evidence \\
{\bf 3C309.1} & 0.905 &    QSO     & 12.75   &   250 & 32 & certain \\
   3C319 & 0.192 &  LERG/FRII & 11.25   &  8000 &128 & likely \\
{\bf 3C321} & 0.096 &  NLRG/FRII & 11.5    &  2000 & 16 & certain \\
{\bf 3C324} & 1.206 &  NLRG/FRII & 12.25   &  4000 &  8 & certain \\
{\bf 3C330} & 0.550 &  NLRG/FRII & 11.75   &   125 & 32 & certain \\
   3C332 & 0.152 &  BLRG/FRII & 11.25   & 16000 &  2 & certain \\
   3C336 & 0.927 &      QSO   & 12      &   250 &  4 & certain \\
{\bf 3C338} & 0.029 &   NLRG/FRI & 10.25   & 16000 & 16 & certain \\
   3C341 & 0.448 &  NLRG/FRII & 11.75   &   125 &  2 & certain \\
   3C343 & 0.988 &    QSO     & 11.75   &  2000 & 16 & possible \\
   3C345 & 0.593 &      QSO   & 13      & 16000 & 64 & certain \\
   3C346 & 0.161 &  NLRG/FRII & 11      & 16000 & 16 & likely \\
   3C349 & 0.205 &  NLRG/FRII & 11.25   & 16000 & 32 & certain \\
{\bf 3C351} & 0.372 &      QSO   & 12.25   &   500 & 32 & certain \\
{\bf 3C356} & 1.079 &  NLRG/FRII & 12.25   &  4000 & 64 & certain \\
   3C371 & 0.051 &   BL Lac   & $<$ 11.5& 16000 & 32 & no evidence \\
 3C379.1 & 0.256 &       RG   & 11.5    &  4000 & 64 & certain \\
{\bf 3C380} & 0.691 &    QSO     & 12.5    &  1000 & 32 & certain \\
{\bf 3C381} & 0.159 &  BLRG/FRII & 11.25   &   250 & 32 & certain \\
{\bf 3C382} & 0.058 &  BLRG/FRII & 11      &   125 &  2 & certain \\
&     & & & & & \\
\hline
\end{tabular}
\end{center}
\end{table*}

\setcounter{table}{3}
\begin{table*}[htb]
\caption{continued}
\begin{center}
\begin{tabular}{l c l r r r l}
\hline
\hline
&     & & & & & \\
Name & z  & Type   & log(L$_{\rm dust}$)& $R$ & $A_{\rm V}$ & Dust ? \\
&     & &  [\Lsun] &[pc] &[mag] & \\
&     & & & & & \\
\hline
   3C386 & 0.017 &   LERG/FRI &  9      &   125 & 16 & possible \\
   3C388 & 0.090 &  LERG/FRII & $<$10.75 &  8000& 64 & no evidence \\
{\bf 3C390.3} & 0.056 &  BLRG/FRII & 11      &   125 &  4 & certain \\
   3C402 & 0.023 &     NLRG   & 10      & 16000 &128 & certain \\
   3C410 & 0.249 &       RG   & 12.75   & 16000 &128 & certain \\
   3C418 & 1.686 &      QSO   & $<$ 8.5 & 16000 &128 & no evidence \\
   3C430 & 0.054 &     NLRG   & $<$ 9   &   250 &128 & no evidence \\
   3C442 & 0.026 &   LERG/FRI & $<$ 8.5 & 16000 &128 & no evidence \\
{\bf 3C445} & 0.056 &  BLRG/FRII & 11.5    &   125 & 16 & certain \\
   3C449 & 0.017 &     BLRG   &  9.75   &  8000 & 64 & possible \\
   3C452 & 0.081 &  NLRG/FRII & 11      &   250 & 64 & certain \\
   3C456 & 0.233 &     LERG   & 11.5    &   250 & 64 & certain \\
 {\bf 3C459} & 0.200 &  NLRG/FRII & 12.25   &  2000 &128 &   certain \\
   3C465 & 0.030 &   LERG/FRI & 10.25   &   125 & 32 &  certain \\
 3C469.1 & 1.336 & LERG/FRII  & $<$ 8.5 & 16000 &128 & no evidence \\
4C+72.26 & 3.532 & RG         & 13.5    & 250   & 64 & no evidence \\ 
\hline
\end{tabular}
\end{center}
\end{table*}

In Fig.~\ref{agn_lrad.ps} extinction and luminosity are held constant
and the influence of the radius, $R$, on the SED is illustrated. We
present models with high extinction $A_{\rm V} = 32$\,mag and
luminosity $L = 10^{12}$ \Lsun \/ as well as moderate extinction
$A_{\rm V} = 8$\,mag and luminosity $L = 10^{11}$ \Lsun, respectively.
When the radius is large, most of the dust is farther away from the
AGN and therefore on average much cooler when compared to models with
small $R$. When the radius of the model is increased while the
extinction is kept fixed more dust is needed for the same
submillimeter flux. As the luminosity is constant the MIR emission
becomes fainter for models with large $R$ and the SED peak is shifted
to longer wavelengths.

Another important feature of the SED is the strength of the PAH
emission. In Fig.~\ref{agn_lrad.ps} (bottom) one notices that PAH band
emission increases for models with increasing cold dust, respectively
large radii.  If $R$ is small the models predict more hot dust and
relatively weak or fully absent PAH bands. This is because most of the
dust emission originates nearby the central source and only few PAH
survive the harsh AGN environment.

To constrain the models one needs MIR and FIR data. The FIR
observations shall give a measure of the SED peak emission.  We have
studied the degeneration of our AGN models. As may be obvious, for
sources where the FIR peak is reasonably well known but there is only
a single data point in the MIR it is still possible to fit the SED by
a combination of low extinction and large radius or vice versa.
However, when more MIR data are available the degeneration almost
disappears.  In particular for galaxies where the ISOCAM data are
suited to indicate the depth of the silicate band the models are quite
unique. This is because the depth of the silicate absorption is
directly related to the visual extinction.

\subsection{Fit to Data}

We selected models from the AGN library (see Sect.~\ref{sec.modpara})
which are consistent with the data. First we require that the models
smoothed to the ISOCAM bandwidths predict the measured ISOCAM fluxes
to within 10\% accuracy. Second, we rejected models which violate any
$3\sigma$ upper limit.  Of the remaining models, we choose the ones
which best fit the optical to FIR/submillimeter data in a least square
sense.  The optical data can be reasonably approximated by a
blackbody of a few thousand K.  We therefore assume that they are due
to (weakly reddened) stars and this stellar component is fit
separately.  As a result one obtains the stellar luminosity (see
Fig.~8).  In 16 out of 69 cases the $A_{\rm V}$ of the model is
$\simless 2$\,mag and then the AGN may also contribute at optical
wavelengths.  The model flux (AGN) is then subtracted from the
measured flux and only the rest is used to evaluate the luminosity of
the stars.  The models are shown as solid lines in
Fig.~\ref{sed.fig}. They fit generally well, exceptions are: 3C273, or
3C382 (see Appendix~\ref{fits.ap} for notes on individual fits).

The results of our modelling procedure are summarised in
Tab.~\ref{tab.dust}. We specify the name, redshift (as available in
NED), type (Veron-Cetty \& Veron 2000, Spinrad et al.~1985), infrared
dust luminosity (Sect.~\ref{sec.modelfit}) and specify a likelihood of
dust detection. We distinguish four cases:

\begin {itemize}
\item {\it No evidence} for dust emission is found.  ISOCAM and
other flux measurements are
consistent with a photospheric or synchrotron component.

\item Detection of dust emission is {\it possible}. ISOCAM photometry
is only slightly above the photospheric component and there are only
far infrared data close to their detection or confusion limit.

\item Detection of dust emission is {\it likely}. Here ISOCAM data are
well above the predicted photospheric component but there is a
possible synchrotron contribution. There are no firm far infrared
detections.

\item Detection of dust emission is {\it certain}. There are reliable
infrared detections well above the predicted photospheric component
and the contribution by synchrotron or free--free emission can be
ruled out.\\

\end {itemize}

\section{Discussion}

\subsection{Energy budget: nucleus versus host}

First, we compare how the AGN and the stars of the host galaxy
contribute to the dust heating.  The influence of a central starburst
is not considered.  One of the three basic parameters of each SED
model is the AGN luminosity.  In our spherical models, which are
all optically thick ($A_{\rm V} \simgreat 1$), the AGN luminosity is
entirely re-radiated by dust in the infrared and it is denoted $L_{\rm
dust}$.  The upper panel of Fig.~\ref{ll} shows the AGN luminosity of
all galaxies detected by ISOCAM as a function of redshift.  In the
lower panel, we divided $L_{\rm dust}$ by the luminosity of all the
stars, $L_{\rm star}$, in the galaxy.  The latter was estimated from
the black--body fits of the photospheric component displayed in
Fig.~\ref{sed.fig} and are not de-reddened.

One notes a trend of $L_{\rm dust}$ with $z$ which is typical for flux
limited samples.  As demonstrated in the lower panel of Fig.~\ref{ll},
the AGN luminosity is much higher than the stellar luminosity, $L_{\rm
star}$, which is concentrated in the near infrared and optical region.
The powerful AGN heats the dust not only in its vicinity but
throughout the galaxy and acceptable model fits to the observed SEDs
were possible only when dust at large (several kpc) distances was
included.

\begin{figure} 
\centerline{\hspace{1.cm} \psfig{file=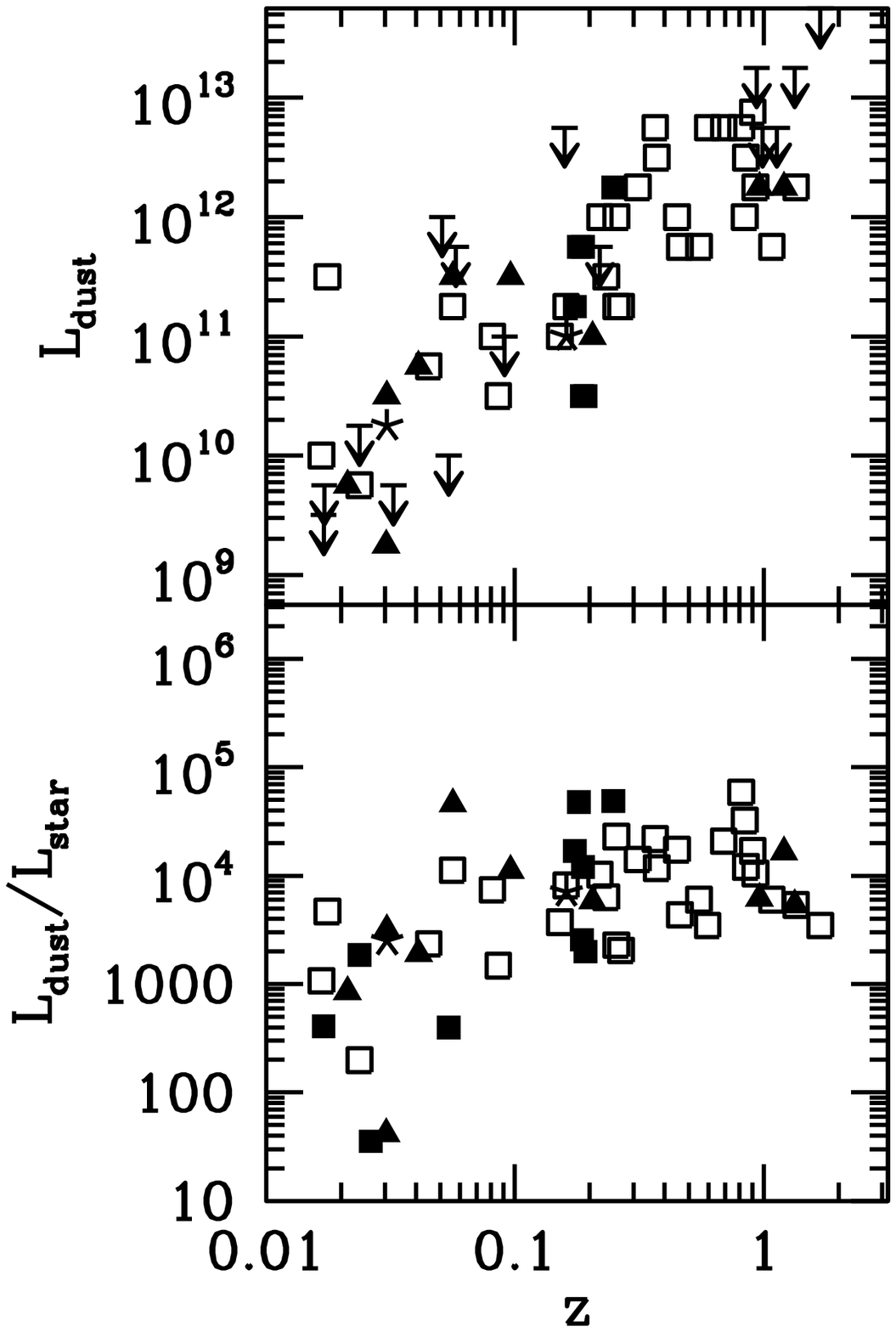,width=12.75cm,height=15cm}} 
\caption{Top: The intrinsic AGN luminosity of the model,
$L_{dust}$,  as function of redshift. Solid symbols are type~1 AGNs:
triangles QSOs and squares BLRGs.  Open squares are NLRGs and upper
limits as of Tab.~\ref{tab.dust}.  Bottom: Luminosity ratio of 
AGN and stellar components as a function of redshift. Symbols as in
top panel.}
\label {ll}
\end{figure}

\begin{figure} [htb]
\centerline{\psfig{file=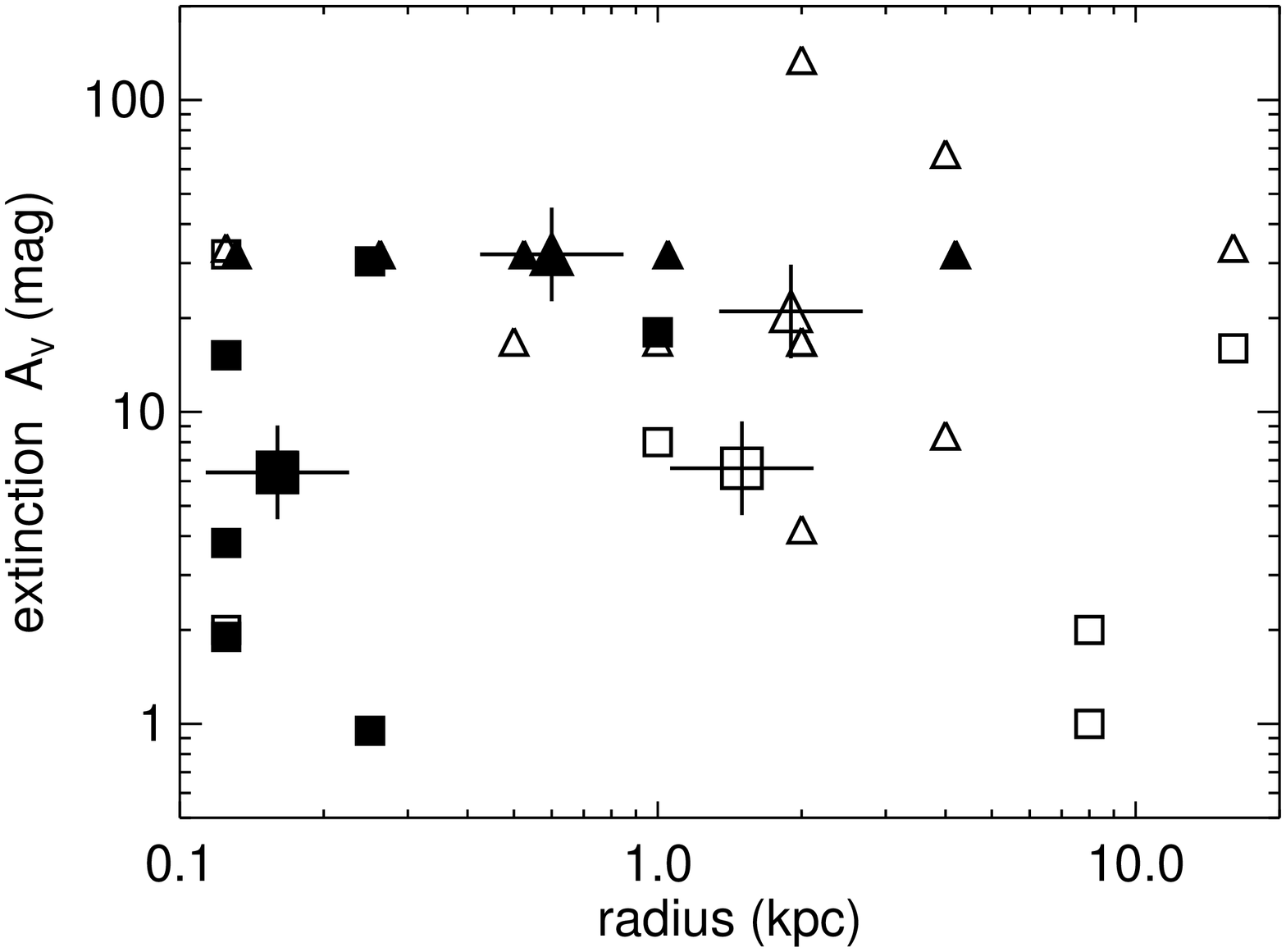,width=7.cm,angle=90}}
\caption{ The best fit model parameters, radius and dust extinction,
for 27 galaxies with well determined SEDs.  The objects are grouped
according to their luminosity (high or low) and spectral type (1 or
2).  There are QSOs (type 1, high $L$, filled triangles), BLRGs (type
1, low $L$, filled squares), NLRGs (type 2,high $L$, open triangles)
and NLRGs (type 2, low $L$, open squares).  For each type, logarithmic
means are marked by crosses.  }
\label {radav_type}
\end{figure}

\subsection{Dust emission and AGN unification}

For 27 AGNs of our sample, marked bold face in Tab.~\ref{tab.dust},
there exist enough photometric data to constrain the spectral shape of
the dust emission.  For a more detailed comparison, we distinguish
between galaxies of optical spectral type 1 and 2 and between objects
of high ($L > 10^{11}$ \Lsun) and low ($L \leq 10^{11}$ \Lsun)
luminosity $L$.  Altogether we thus have four classes: 6 BLRGs (type
1), 5 QSOs (type 1), 6 low $L$ and 10 high $L$ NLRGs (both of type 2).
According to the unification scheme, type 1 AGNs are viewed nearly
pole--on and have visible broad line region, whereas type 2 AGNs are
viewed edge--on, so that the large extinction from the AGN torus
obscures the broad-line region.

In Fig.\ref{radav_type}, we plot for the 27 objects the best fit
parameters: radius and visual extinction.  AGNs of type 1 occupy the
left and those of type 2 the right part of the frame, so the models
for them differ foremost by the size parameter or radius $R$.  This
was already evident from the infrared SEDs in Fig.~\ref{agn_lrad.ps}
(bottom) where small model radii correspond to BLRGs and large radii
to low luminosity NLRGs.  For low luminosities, both NLRGs and BLRGs
are fit by moderate extinction ($\overline{A_{\rm V}} \sim 7$\,mag).
For high luminosity NLRGs and QSOs, we find higher values of $A_{\rm
V}$, the averages are $\overline{A_{\rm V}} = 20$ and 30\,mag,
respectively.  This trend is reflected for QSOs and high luminosity
NLRGs in the top panel of Fig.~\ref{agn_lrad.ps}, where SEDs with a
radius close to the mean value of both AGN types can be compared.

In summary, the model spectra of BLRGs and QSOs
(Fig.~\ref{agn_lrad.ps}, bottom) peak at $\sim 40\mu$m, whereas high
and low luminosity NLRGs attain their maxima much longer wavelengths,
around 150 $\mu$m.  Therefore, the dust in AGNs of type 1 is warmer
than in those of type 2.  This difference is qualitatively consistent
with the unification schemes.  For type~1 AGNs, which are viewed close
to pole--on, the broad--line region is visible and unobscured hot dust
dominates the SED.  By contrast, towards NLRG, which are viewed
face--on, large extinction from the torus obscures the broad--line
region and pre-dominantly cooler dust is detected.  The type 1 AGNs
also show much less PAH emission than type 2.  The weakness of the
band emission in type 1, compared to type 2, is probably due to PAH
destruction.  Averages of observed SED are presented by Freudling et
al. (2003) confirming the trends indicated in the models presented
here.

\section{Conclusions}

ISOCAM detected 80\% (71 out of 88) 3CR sources.  We find evidence of
hot dust in 53 radio galaxies or 75\% of the sources detected in our
survey.  For each detected source, we compiled photometry from optical
to radio wavelengths.

The extrapolation of radio core fluxes to ISOCAM wavelengths shows
that synchrotron emission from the core is usually negligible in the
MIR, exceptions being flat spectrum radio sources.  The contribution
of the host galaxy to the MIR is also generally small, here exceptions
are a few sources of type FRI.  Thus for most objects the origin of
the MIR can be attributed mainly to dust.  This dust is hot and heated
by the central engine.  The emission comes from large grains with
temperatures of a few hundred Kelvin and small grains and PAHs
undergoing temperature fluctuations.  This picture is supported by
simple radiative transfer calculations.  Most SEDs can be successfully
fit between 1 and 1000$\mu$m by a three parameter model where AGN
luminosity, size of the galaxy and extinction are varied. Our
modelling shows that the broad band data are consistent with AGN only
heating of the dust.

For similar luminosities, we compare AGNs of type 1 and 2.  We find
that the model parameters depend strongly on the AGN type and a
dichotomy of the infrared SEDs is derived.  The IR fluxes of BLRGs and
QSOs peak around 40$\mu$m while the maximum of the dust emission for
NLRGs is reached at $\simgreat 100\mu$m.  Therefore, the dust in type
1 AGNs is warmer than in type 2.  The models also predict much weaker
PAH emission in type 1 AGNs than in type 2, as a result of
evaporation.  Unfortunately, PAH bands cannot be resolved by the
broad--band ISOCAM observations.

AGNs and starbursts are accompanied by tremendous IR luminosities.
Although the possible coexistence of both energy sources and their
relative contributions to the dust heating is difficult to assess,
starbursts tend to favour PAH emission and this may serve as a possible
discriminator of activity.

\acknowledgements

We thank the referee, Cristina Popescu, for constructive and
detailed suggestions. Rosario Lorente at the ISO Data Centre for
archiving support.  Martin Haas thanks for grants from the
Nordrhein-Westf\"alische Akademie der Wissenschaften, funded by the
Federal State Nordrhein-Westfalen and the Federal Republic of
Germany. This research has made use of the SIMBAD database, operated
at CDS, Strasbourg, France and the NASA/IPAC Infrared Science Archive
and Extragalactic Database (NED) which is operated by the Jet
Propulsion Laboratory, California Institute of Technology, under
contract with the National Aeronautics and Space Administration. This
publication makes use of data products from the Two Micron All Sky
Survey, which is a joint project of the University of Massachusetts
and the Infrared Processing and Analysis Center/California Institute
of Technology, funded by the National Aeronautics and Space
Administration and the National Science Foundation. CIA is a joint
development by the ESA Astrophysics Division and the ISOCAM
Consortium. The ISOCAM Consortium is led by the ISOCAM PI,
C. Cesarsky.

\appendix
\section{Notes on individual galaxies}\label{galnotes.ap}

Our remarks below concern the following points: (1) previous
indications of dust, (2) X-ray emission, (3) jets, (4) high excitation
line emission, (5) morphology, (6) companions.

\noindent 
{\it 3C002} ($z = 1.037$): Fuzzy optical structure similar to radio morphology
(Ridgway \& Stockton 1997).

\noindent 
{\it 3C006.1} ($z = 0.804$): There is an optical extension with position angle
roughly equal to that of the radio axis (McCarthy et al.~1997).

\noindent {\it 3C013} ($z = 1.351$):  There are two faint tail-like
structures to the south and east of the nucleus. The core radio source is a
double with a size of 28$''$.

\noindent {\it 3C017} ($z = 0.2$): X-ray source (Siebert et al.~1996)
with a strong and broad H${\alpha}$ line (Eracleous \& Halpern 1994).

\noindent {\it 3C018} ($z = 0.19$): Shows a rich emission line spectrum with
extremely high ionization states (Tadhunter et al.~1993).

\noindent 
{\it 3C020} ($z = 0.17$): Two cones of emission and a possibly dust disk are
detected by de Koff et al. (2000). The large--scale FR~II class radio structure
is aligned roughly perpendicular to the proposed dust lane.

\noindent {\it 3C022} ($z = 0.935$): From JKL' images. Simpson et
al.~(1999) conclude that 80\% of the light in a 3$''$ aperture is from
a weakly reddened ($A_{\rm V} =2-5$mag) quasar; extended near IR light
is presumably photospheric.

\noindent {\it 3C031} ($z = 0.0167$): Associated with NGC383. HST images show a
face--on dust disk with a diameter of 7$''$ and an unresolved core (Xu et
al.~2000). Martel et al. (1999) confirm an optical synchrotron jet. The
molecular gas content is $M(\rm{H}_2) \sim$ 10$^9$ \Msun\ (Leon et al.~2001).

\noindent 
{\it 3C33.1} ($z = 0.181$): Broad line radio galaxy in a region of high
foreground stellar density (McCarthy et al.~1995). A star is superimposed
southeast of the nucleus.

\noindent 
{\it 3C048} ($z = 0.367$): Significant part of the OIII line emission is
confined to the 0.5$''$ central region (Axon et al.~2000).

\noindent {\it 3C61.1} ($z = 0.186$): Narrow-line galaxy.

\noindent {\it 3C66B} ($z = 0.0212$): HST observations confirm an
optical synchrotron jet (Martel et al.~1999).  Orbital motion in this
radio galaxy provides evidence for a supermassive black hole binary
(Sudou et al 2003).

\noindent 
{\it 3C071} (NGC1068, $z = 0.003$): Famous Seyfert~2; MIR images of high
spatial resolution are obtained by Siebenmorgen et al. (2004).

\noindent {\it 3C076.1} ($z = 0.0325$): A galaxy with no optical or near IR
features (Colbert et al.~2001). 

\noindent {\it 3C079} ($z = 0.256$): Narrow H${\alpha}$ and H${\beta}$ lines
(Laing et al.~1994).

\noindent 
{\it 3C083.1} (NGC1265, $z = 0.025$): Edge--on radio galaxy with radio jet and
nearly orthogonal narrow dust lane of $\sim$ 1.5\,kpc. There is a bright
foreground star near the center (Martel et al.~1999).

\noindent {\it 3C084} (NGC1275, $z = 0.0176$): Extensively studied
giant elliptical galaxy. It has filamentary dust structures which
extend out to 17\,kpc (de Koff et al.~2000).

\noindent 
{\it 3C098} ($z = 0.031$): Seyfert~2 with extended X--ray emission and
a compact central X--ray source (Hardcastle \& Worrall 1999). No
optical/near IR features (Colbert et al.~2001).

\noindent 
{\it 3C231} (M82, $z = 0.001$): Archetype starburst galaxy; see Kr\"ugel \&
Siebenmorgen (1994) for a model of the dust emission.

\noindent {\it 3C249.1} (PG~1100+772, $z = 0.312$): Absorption line quasar.

\noindent {\it 3C264} ($z = 0.022$): An apparent optical ``ring'' is claimed to
be produced by a circumnuclear dust disk (Baum et al.~1997).

\noindent {\it 3C265} ($z = 0.811$): Complex optical structure without a point
source at the nucleus. Probably highly obscured AGN ($A_{\rm V} \ge
15$mag, Simpson et al.~1999).

\noindent 
{\it 3C270} (NGC4261, $z = 0.007$): Quasar with nuclear dust disk
perpendicular to the large-scale radio jets (Jaffe et al.~1996).
Indication of vigorous star formation around the AGN (Canalizo \&
Stockton 2000).  The quasar suffers little extinction (Leon et
al.~2001).

\noindent {\it 3C272.1} (M84, $z = 0.004$): Elliptical galaxy with dust
extending out to almost 1~kpc (de Koff et al.~2000).

\noindent 
{\it 3C273} ($z = 0.158$): First discovered QSO (Schmidt 1963).

\noindent 
{\it 3C274} (M87, NGC4486, $z = 0.004$): It has an ionized disk surrounding the
central black hole (Ford et al.~1994, Harms et al. (1994) of $2.6 \cdot 10^9$
\Msun\ (Lauer et al.~1992).

\noindent {\it 3C277.3} (Coma~A, $z = 0.086$): Elliptical with radio
jet.

\noindent {\it 3C286} ($z = 0.849$): Contains high excitation line emission
(Boisse et al.~1998).

\noindent {\it 3C288.1} ($z = 0.961$): Double radio quasar (Akujor et
al.~1994).

\noindent {\it 3C293} (UGC~08782, $z = 0.045$): Has a double nuclei core with a
steep radio spectrum and filamentary dust lanes (Martel et al.~1999).

\noindent {\it 3C295} ($z = 0.464$): Extended double-lobe radio galaxy
without a core (Akujor et al.~1994, Fomalont et al.~2000).

\noindent {\it 3C296} (NGC5532, $z = 0.0237$): This elliptical galaxy hosts a
sub-kiloparsec, elongated and sharp dusty structure centerd on its
nucleus (Martel et al.~1999).

\noindent {\it 3C303.1} ($z = 0.267$): The nucleus has an elongated shape. The
extended radio source is an asymmetric double $\sim$ 1.8$''$ in size,
elongated perpendicular to two dust lanes (de Koff et al.~2000).

\noindent {\it 3C305} ($z = 0.039$): An inclined disc of gas and dust
encircling a quasar reddened by $A_{\rm V} > 4$ has been observed by Jackson
et al. (2003).

\noindent {\it 3C305.1} ($z = 1.132$): The NED coordinates have an
uncertainty of 4$''$. In the DSS, a contaminating source is visible
8$''$ away. The radio emission of this high-redshift galaxy forms a
core--free structure with two lobes separated by $\sim 2''$.

\noindent {\it 3C309.1} ([HB89]~1458+718, $z = 0.905$): This quasar has a
compact steep radio spectrum (Peng et al.~2000) and remains unresolved
in the optical (de Vries et al. 1997).

\noindent {\it 3C319} ($z = 0.192$): Resembles a normal
non-star-forming elliptical with an optical long axis almost
perpendicular to the radio axis (Roche \& Eales 2000).

\noindent {\it 3C321} ($z = 0.096$): Blue, high surface brightness galaxy. The
disturbed dust lane with a projected length of $\sim$ 3.7$''$ breaks in
two components belonging to two merging galaxies (Roche \& Eales
2000).

\noindent 
{\it 3C324} ($z = 1.206$): A gravitationally--lensed, extremely elongated high
redshift radio galaxy (Chambers et al.~1996) with a steep radio spectrum
($\alpha \sim$ 1.07, Brunetti et al.~1997). It is a ROSAT source with an X--ray
luminosity of $4.4 \times 10^{10}$ \Lsun \/ (Hardcastle \& Worrall 1999).  From
SCUBA and Effelsberg observations Best et al. (1998) conclude that the
submillimeter signal is not associated with synchrotron emission.

\noindent {\it 3C330} ($z = 0.550$): Optically very luminous and relatively
blue galaxy. The disk structure is dominated by line emission
(McCarthy et al.~1996).

\noindent {\it 3C332} ($z = 0.152$): This broad line radio galaxy shows in HST
images a smooth elliptical nuclear region (Taylor et al.~1996).

\noindent {\it 3C336} (PKS~1622+23, $z = 0.927$): This source has a large
angular radio size (28$''$).  The brightest ISOCAM pixel is close to
the radio source FIRST~J162438.8+234505. The latter may be a resolved
component of PKS~1622+23.

\noindent 
{\it 3C338} (NGC6166, $z = 0.030$): Elliptical in the cluster Abell~2199.  An
arc-like dust feature extends for $\sim 3''$ from the nucleus to the West
(Capetti et al.~2000).

\noindent 
{\it 3C341} ($z = 0.448$): As the NED coordinates are uncertain by 3$''$, we use
coordinates derived from DSS2. Martel et al. (1999) confirm an optical
synchrotron jet.

\noindent {\it 3C343} ($z = 0.988$): An optically elongated and not very
centrally concentrated quasar. The optical and radio emissions may
have similar shapes (de Vries et al. 1997).

\noindent {\it 3C343.1} ($z = 0.75$): No radio core has been detected (Ludke et
al.~1998).

\noindent {\it 3C345} ($z = 0.595$): A superluminal optically violently
variable quasar (Penston \& Cannon 1970, Katajainen et al.~2000). The
spectrum is flat for wavelengths $>$ 3mm and steepens for $\lambda
\leq 3$\,mm (Rantakyroe et al.~1998). As NED coordinates are uncertain
by 3$''$ we use for astrometry the VLA map by Leahy \& Perley (1991).

\noindent {\it 3C346} ($z = 0.162$): Narrow-line radio galaxy showing both
extended and compact X-ray emission (Hardcastle \& Worrall 1999). In
the optical there is a jet and a bright core (de Koff et al.~1996).
We ignore deviant fluxes by Morabito et al. (1982).

\noindent 
{\it 3C349} ($z = 0.205$): There are two companions, a faint filament suggests
an interaction within the triple (Roche \& Eales 2000).

\noindent {\it 3C351} ($z = 0.372$): Radio-loud QSO of moderate X-ray
luminosity. The spectrum from 0.3--10.1$\mu$m can be fitted with a
broken power law (Neugebauer et al.~1987).

\noindent {\it 3C356} ($z = 1.079$): In the optical, there are two nuclei 5$''$
apart (McCarthy et al.~1995, 1997) and both are detected in the radio
(Eales \& Rawlings 1990, Fernini et al.~1993).  We identify the one
which is brighter in the K band with the nucleus of the radio source.
3C356 is detected at X--rays with ROSAT (Hardcastle \& Worrall 1999,
Crawford \& Fabian 1996).

\noindent {\it 3C368} ($z = 1.131$): The spectrum from radio to optical
wavelengths can be fit by a power-law with a break in the infrared.
There are two knots emitting synchrotron emission in the optical
(O'Dea et al.~1999). Near infrared spectroscopy by Jackson \& Rawlings
(1997) shows that the nucleus is contaminated by a galactic M star.

\noindent {\it 3C371} ($z = 0.051$): BL Lac object with an unresolved
HST point source (Martel et al.~1999).

\noindent {\it 3C379.1} ($z = 0.256$): Shows extended OIII emission over a
scale of 9$''$ (McCarthy et al.~1995).

\noindent {\it 3C380} ($z = 0.692$): There are two optical hot spots coinciding
with radio peaks (Ludke et al.~1998). Therefore, the optical hot spots
probably originate from synchrotron radiation (de Vries et al. 1997).

\noindent {\it 3C381} ($z = 0.161$): Shows extended OIII emission on a scale of
$\sim$ 20$''$ (McCarthy et al.~1995). There is a low surface
brightness companion 5.5$''$ away (Roche \& Eales 2000).

\noindent {\it 3C382} ($z = 0.058$): A classic broad-lined radio galaxy whose
optical luminosity is dominated by an extremely blue AGN (Roche \&
Eales 2000).

\noindent {\it 3C386} ($z = 0.017$): An elliptical with a compact
optical nucleus.  There are two point sources of similar magnitude
$\sim 3''$ away from the nucleus.  They might be faint foreground
stars or small companion galaxies (Martel et al.~1999).

\noindent {\it 3C388} ($z = 0.091$): This radio galaxy has an unresolved
optical nucleus (Martel et al.~1999) and two companions with a nuclear
separation of 7$''$ (Roche \& Eales 2000).  The X--ray images show
significant diffuse emission (Hardcastle \& Worrall 1999).

\noindent 
{\it 3C390.3} ($z = 0.056$): A double-lobed, broad line blazar (Fomalont et
al.~2000) detected also at X--rays (Hardcastle \& Worrall 1999).  The optical
band is dominated by a nuclear point source (Martel et al.~1999).

\noindent {\it 3C401} ($z = 0.201$): A radio galaxy with a double nucleus
(separation 3.6$''$) and several neighbours (Roche \& Eales 2000). 

\noindent {\it 3C402} ($z = 0.025$): An elliptical galaxy with an FR~I
morphology and a double radio nucleus. There is a massive nuclear disk
of gas and dust (De Juan et al.~1996).

\noindent {\it 3C410} ($z = 0.249$): An elliptical with an optical jet
candidate (de Koff et al.~1996).

\noindent {\it 3C418} ($z = 1.69$): Has several nearby but faint companions
(Lehnert et al.~1999).

\noindent 
{\it 3C430} ($z = 0.056$): An elliptical with a narrow, arc-shaped dust lane
that slices the core (Martel et al.~1999).

\noindent {\it 3C433} ($z = 0.101$): For the near infrared we use the data of
2MASXi~J2123445+250427. The absorption map by de Koff et al. (2000)
shows patchy dust features.

\noindent {\it 3C442} ($z = 0.026$): A galaxy pair interacting with
NGC7237 (De Vaucouleurs et al.~1976).

\noindent {\it 3C445} ($z = 0.056$): A broad emission line object (Eracleous \&
Halpern 1994) with an unresolved nucleus (Martel et al.~1999).

\noindent {\it 3C449} ($z = 0.017$): A low--power radio galaxy with multiple
X--ray components and a nearby companion (Worrall \& Birkinshaw
2000). Leon et al. (2001) derived from CO a molecular gas mass of
$M(\rm{H}_2) \sim$ 2.3$\times 10^8$ \Msun. The HST images reveals a
5$''$ inner disk--like feature (Zirbel \& Baum 1998).

\noindent {\it 3C452} ($z = 0.081$): de Koff et al. (2000) find indication of a
faint dust lane near the nucleus. The radio source has a bright core
and two lobes.

\noindent {\it 3C454.1} ($z = 1.841$): This source appears in the radio as a
double with no core (Ludke et al.~1998).  In the optical continuum one
sees two nuclear condensations separated by 2$''$ (Djorgovski et
al.~1988). 

\noindent 
{\it 3C456} ($z = 0.233$): The radio core of this FR~II galaxy lies
within 1$''$ of the optical core position (Harvanek \& Hardcastle
1998). The near IR shows extended emission (de Vries et al.~1997).

\noindent 
{\it 3C459} ($z = 0.02$): This steep spectrum ULIRG (IRAS 23140+0348)
shows optical absorption features suggestive of young stellar
populations (Miller 1981) and broad wings on both H$\alpha$ and
H$\beta$ (Yee \& Oke 1978). In the optical there are two polarisation
components (Draper et al. 1993) none of which coincide with the radio
axis. The radio emission (Condon et al. 1998) is extended to only
$\sim 10''$ (48kpc).

\noindent {\it 3C465} (NGC7720, $z = 0.030$): HST reveals a dusty
kpc-size disk (Martel et al.~1999) which is roughly perpendicular to
the radio jets (de Koff et al.~2000).

\noindent {\it 3C469.1} ($z = 1.336$): There is a double nucleus in the optical
not aligned with the radio axis (McCarthy et al.~1997).

\noindent 
{\it 4C+72.26} ($z = 3.536$): The galaxy with the greatest red shift in our sample.

\noindent 
{\it NGC5532NED02} ($z \sim$ 0.0237): As no redshift is found in NED
we use the value for NGC5532 (3C296).

\noindent 
{\it NGC7236} ($z = 0.026$): A peculiar radio galaxy merging with
NGC7237. It is also detected at X--ray (Hardcastle \& Worrall 1999).

\noindent 
{\it PMN~J0214-1158} ($z = 2.34$): A large high redshift radio source
with a steep radio spectrum; simple double morphology and jet
signatures (Roettgering et al.~1994).

\noindent 
{\it IRAS~F~17130+5021} ($z = 0.05$): A galaxy pair 15$''$ away
associated with 2MASXi~J~1714167+501816.

\section{Notes on individual fits}\label{fits.ap}

\noindent {\it 3C006.1}: Evidence for dust comes from ISOCAM 12$\mu$m
(this work) and weak millimeter detection by Haas et
al. (2001). Evidence for hot dust comes from ISOCAM only.

\noindent {\it 3C013 and 3C017}: Fitting the visible and ISOCAM data
points by a Planck function requires a rest frame temperature of
2600\,K.  If one allows for moderate reddening of $A_{\rm V} = 1$\,mag
the temperature would rise to 2900\,K. Such a value is still lower
than the expected mean stellar temperature in the galaxy and therefore
the presence of hot dust is likely.

\noindent {\it 3C018}: 

\noindent {\it 3C018}: IRAS fluxes suggests A$_{\rm{V}}$ larger than
30mag. Evidence for hot dust from ISOCAM only.

\noindent {\it 3C020}: Evidence for hot dust by ISOCAM.

\noindent {\it 3C022}: Fitting the visible and ISOCAM data points by a
Planck function requires a rest frame temperature of 3000\,K.  If one
allows for moderate reddening of $A_{\rm V} = 1$\,mag in the galaxy
the temperature would rise to 3600\,K. Such a value is compatible to a
normal stellar population. However, near IR imaging (Simpson et
al.~1999) suggests a possible detection of dust.

\noindent {\it 3C031}: ISOCAM emission comes from a compact region of about
8$''$ whereas the shown optical data are the magnitudes of the whole
galaxy (180$''$).  The MIR emission of the stars are therfore only
a small contribution to the ISOCAM fluxes.  Evidence for hot dust by ISOCAM.

\noindent {\it 3C033.1, 3C048 and 3C061.1}: Evidence for dust by
ISOCAM and ISOPHOT.

\noindent {\it 3C66B}: Evidence for hot dust comes from ISOCAM data.

\noindent {\it 3C079}: Evidence for dust from ISOCAM and ISOPHOT.
      
\noindent {\it 3C083.1}: Star superposed just 3$''$ southeast of the
nucleus makes modelling of this galaxy impossible.

\noindent {\it 3C098}: Evidence for hot dust from ISOCAM only.
        
\noindent {\it 3C231}: M82 starburst model as described by Kr\"ugel \&
Siebenmorgen (1994).
      
\noindent {\it 3C249.1}: Evidence for dust from ISOCAM and ISOPHOT.

\noindent {\it 3C265}: Evidence for dust from ISOCAM only.
  
\noindent {\it 3C270}: No evidence of dust from ISOCAM.

\noindent {\it 3C272.1}: Evidence for dust from ISOPHOT only (Haas et
al.~2001).

\noindent {\it 3C273}: Possible hot dust detection by ISOCAM.

\noindent {\it 3C274}: No evidence of dust from ISOCAM.

\noindent {\it 3C277.3}: Evidence for dust from ISOCAM only.

\noindent {\it 3C286}: Evidence for hot dust by ISOCAM only.

\noindent {\it 3C288.1}: IRAS 60 $\mu$m maybe due to confusion.
Likely detection of hot dust by ISOCAM. 

\noindent {\it 3C293}: Evidence for hot dust from ISOCAM only.

\noindent {\it 3C295}: Evidence for dust from ISOCAM and ISOPHOT.

\noindent {\it 3C296}: IRAS 100 $\mu$m maybe due to
confusion. Likely detection of hot dust by ISOCAM. 

\noindent {\it 3C303.1}: 
Fitting the visible and ISOCAM data points by a Planck function
requires a rest frame temperature of 2600\,K.  If one allows for
moderate reddening of $A_{\rm V} = 1$\,mag the temperature would rise
to 2900\,K. Such a value is still lower than the expected mean stellar
temperature in the galaxy. Evidence for dust from ISOCAM and ISOPHOT.

\noindent {\it 3C305}: Evidence for dust by ISOCAM and ISOPHOT.

\noindent {\it 3C309.1}: Evidence for dust by ISOCAM and ISOPHOT.

\noindent {\it 3C319}: Likely detection of hot dust by ISOCAM. 

\noindent {\it 3C321}: Evidence for dust from ISOCAM and ISOPHOT.

\noindent {\it 3C324}: Evidence of hot dust from ISOCAM only.
              
\noindent {\it 3C330}: Evidence for dust from ISOCAM only.

\noindent {\it 3C332}:  Evidence for hot dust from ISOCAM only.            

\noindent {\it 3C336}:  Evidence for hot dust from ISOCAM only.   

\noindent {\it 3C338}: Evidence for dust from IRAS only.

\noindent {\it 3C341}: Evidence for dust from ISOCAM only.  

\noindent {\it 3C343}: Possible detection of dust by ISOCAM only.

\noindent {\it 3C345}: Evidence for hot dust from ISOCAM only.  

\noindent {\it 3C346}: Evidence for dust from ISOCAM only. Large
deviation with ISOPHOT 12 $\mu$m fluxes given by Haas et al. (2001).

\noindent {\it 3C349}: Evidence for dust from ISOCAM only.

\noindent {\it 3C351}:  Evidence for dust from ISOCAM and ISOPHOT.

\noindent {\it 3C356}: Evidence for dust from ISOCAM and ISOPHOT.

\noindent {\it 3C371}: No evidence for dust since the contribution of
the synchrotron radiation to the ISOCAM data is unclear. 

\noindent {\it 3C379.1}: Evidence for dust by ISOCAM  only.

\noindent {\it 3C380}: The measured radio core flux is similar to the
total radio emission, so that the synchrotron component, interpolated to
the IR, is strong. Still the ISOCAM detection gives evidence of dust
emission.

\noindent {\it 3C381}: Evidence for dust from ISOCAM and ISOPHOT.
   
\noindent {\it 3C386}: Evidence for dust from ISOCAM  at 14.3$\mu$m only.

\noindent {\it 3C401}: In the infrared/submillimeter there are only
upper limits which hints to model this galaxy.
             
\noindent {\it 3C402}: ISOCAM emission comes from a compact region of
about 9$''$ whereas the shown optical data are the magnitudes of the
whole galaxy (55$''$).  The MIR emission of the stars is therefore
only a small contribution to the ISOCAM fluxes. Evidence for dust from
ISOPHOT (Haas et al.~2001) only.

\noindent {\it 3C410}: Evidence for dust from ISOCAM only.

\noindent {\it 3C442}: ISOCAM emission comes from a compact region ($<
5''$) whereas the shown optical data are the magnitudes of the whole
galaxy (130$''$).  The MIR emission of the stars is therefore only
a small contribution to the ISOCAM fluxes. Weak 12$\mu$m detection
might be still photospheric. No evidence for dust.

\noindent {\it 3C449}: ISOCAM emission comes from a compact region of about
7$''$ whereas the shown optical data are the magnitudes of the whole
galaxy (90$''$).  The MIR emission of the stars is therefore only a
small contribution to the ISOCAM fluxes. Evidence for dust comes from
ISOCAM only.

\noindent {\it 3C449}: Evidence for hot dust from ISOCAM only.
    
\noindent {\it 3C452}: Evidence for dust comes from ISOCAM only.

\noindent {\it 3C456}: Fitting the visible and ISOCAM data points by a
Planck function requires a rest frame temperature of 2300\,K.  If one
allows for moderate reddening of $A_{\rm V} = 1$\,mag the temperature
would rise to 2600\,K. Such a value is still lower than the expected
mean stellar temperature in the galaxy. Evidence for dust from ISOCAM
only.
    
\noindent {\it 3C459}:  Evidence for dust from ISOCAM and ISOPHOT.

\noindent {\it 3C465}: ISOCAM emission comes from a compact region of
about 10$''$ whereas the shown optical data are the magnitudes of the
whole galaxy (180$''$).  The MIR emission of the stars is therefore
only a small contribution to the ISOCAM fluxes. Evidence for hot dust
from ISOCAM only.

\noindent {\it 3C469.1}: Evidence for dust from ISOCAM and ISOPHOT.
     
\noindent {\it 4C+72.26}: Our highest redshift quasar (z=3.532).  The
photospheric contribution is unknown, so that there is yet no
evidence for dust. The model is not well constrained by the sparse
infrared/submillimeter data.

\clearpage
\begin{figure*}
\label{images}
\center {\epsfig{file=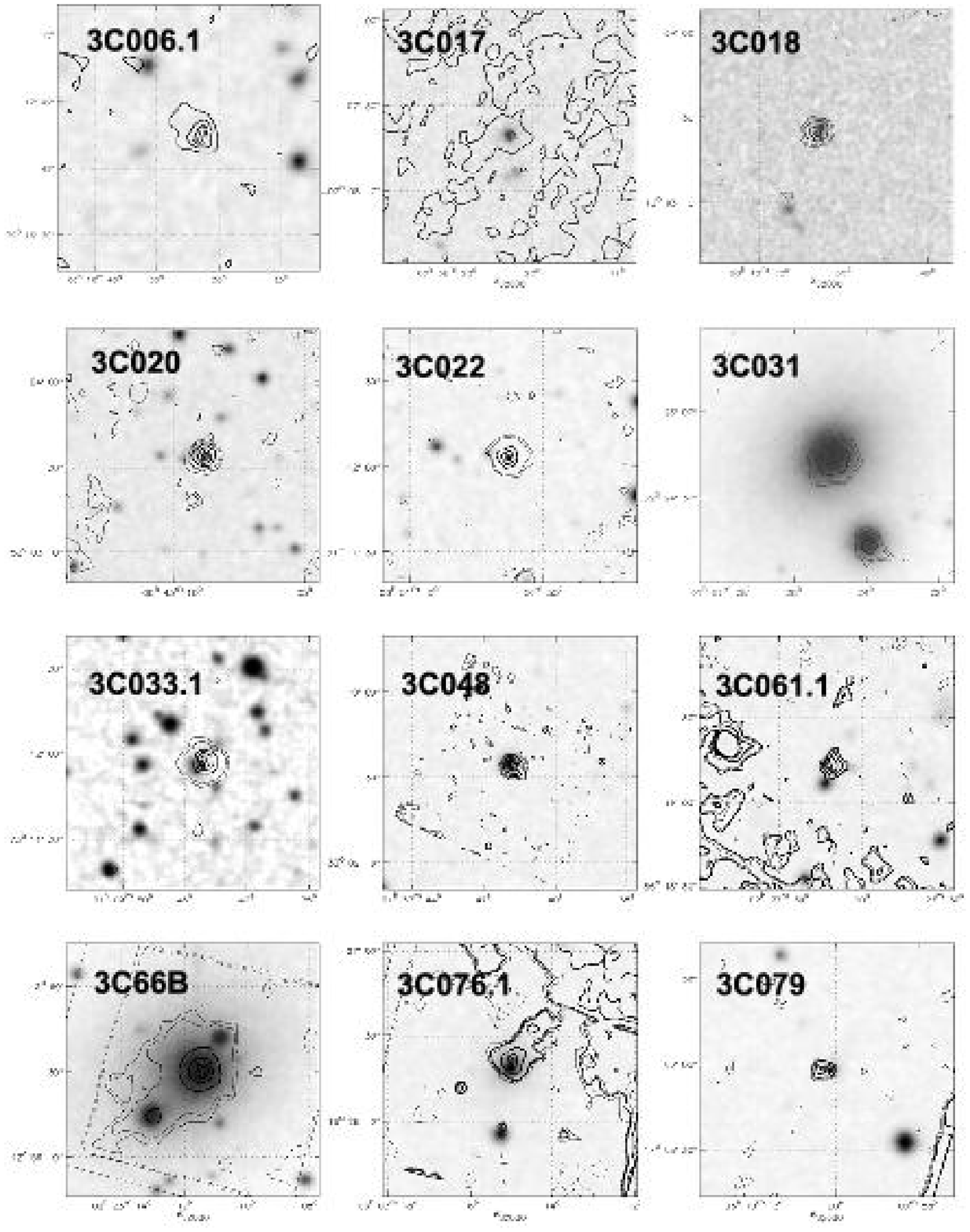}} 
\caption{ISOCAM contours overlayed on DSS grey scale images centered
on NED position of the object marked in bold. For each object we
specify the ISOCAM filter and contour levels in
$\mu$mJy/arcsec$^2$. Negative contours at -3$\sigma$, -$\sigma$ are
shown dashed. {\bf 3C006.1} at 12.0$\mu$m (LW10): $\pm$4.4, $\pm$8.7,
15.7, 102.5, 153.8\,.  {\bf 3C017} at 12.0$\mu$m (LW10): $\pm$ 22.2,
$\pm$44.3, 79.7, 521.9, 782.8. {\bf 3C018} aqt 12.0$\mu$m (LW10):
$\pm$50.9, $\pm$101.7, 152.6, 183.1, 204.0.  {\bf 3C020} at 12.0
$\mu$m (LW10): $\pm$25.1, $\pm$50.2, 90.4, 108.8, 163.2\,.  {\bf
3C022} at 12.0$\mu$m (LW10): $\pm$41.4, $\pm$82.8, 124.2, 149.0,
161.1\,.  {\bf 3C031} at 12.0$\mu$m (LW10): $\pm$65.6, $\pm$131.3,
196.9, 236.3, 260.1\,.  {\bf 3C033.1} at 12.0$\mu$m (LW10): $\pm$77.1,
$\pm$96.4, 154.2, 231.3, 277.6\,.  {\bf 3C048} at 14.3 $\mu$m (LW3):
$\pm$342.5, $\pm$228.4, 342.5, 749.9, 1124.9, 1349.9\,.  {\bf 3C061.1}
at 12.0$\mu$m (LW10): $\pm$ 30, $\pm$ 70, 157, 314, 471, 566\,.  {\bf
3C66B} at 6.7$\mu$m (LW2): $\pm$22.2, $\pm$44.3, 66.5, 79.7, 95.1\,.
{\bf 3C076.1} at 4.5$\mu$m (LW1): $\pm$11.9, $\pm$23.8, 35.7, 42.8,
53.1\,.  {\bf 3C079} at 12.0$\mu$m (LW10): $\pm$89.6, $\pm$106.7,
213.5, 320.2, 384.3\,.  }
\end{figure*}

\clearpage
\setcounter{figure}{0}
\begin{figure*}
\center {\epsfig{file=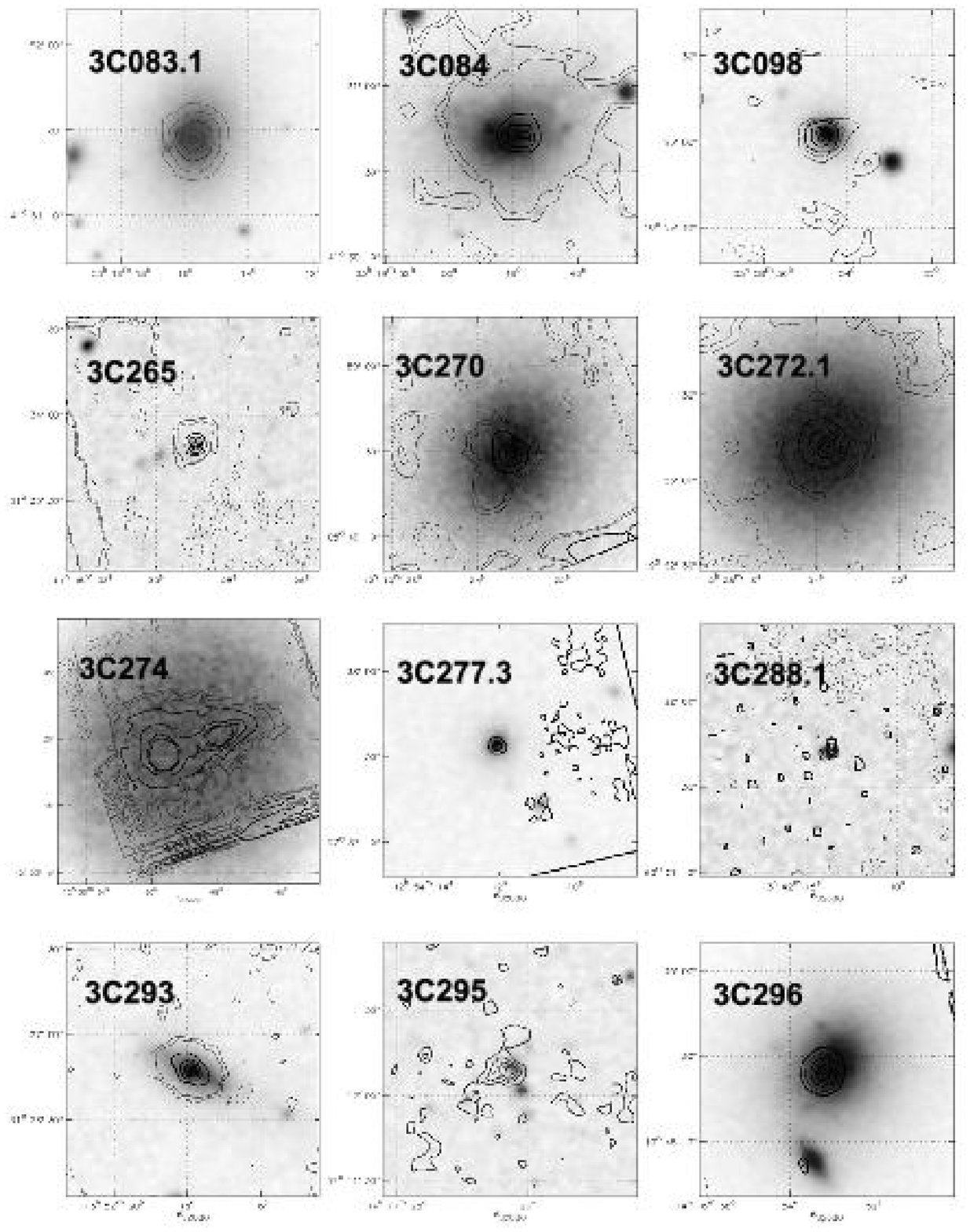}}
\caption{ -- continued --: {\bf 3C079} at 12.0$\mu$m (LW10):
$\pm$89.6, $\pm$106.7, 213.5, 320.2, 384.3\,.  {\bf 3C083.1} at 4.5
$\mu$m (LW1): $\pm$61.1, $\pm$91.6, 227.6, 341.5, 409.8\,.  {\bf
3C084} at 6.7$\mu$m (LW2): $\pm$129.4, $\pm$194.1, 759.9, 1139.9,
1367.9\,.  {\bf 3C098} at 14.3$\mu$m (LW3): $\pm$10.0, $\pm$14.2,
21.3, 30.0, 35.9\,.  {\bf 3C265} at 12.0$\mu$m (LW10): $\pm$34.7,
$\pm$69.4, 124.9, 477.0, 715.6\,.  {\bf 3C270} at 14.3$\mu$m (LW3):
$\pm$27.7, $\pm$38.0, 76.1, 114.1, 137.0\,.  {\bf 3C272.1} at 14.3
$\mu$m (LW3): $\pm$95.2, $\pm$95.2, 142.7, 306.4, 459.5, 551.4\,.
{\bf 3C274 (M87)} at 12.0$\mu$m (LW10): $\pm$55.6, $\pm$111.2, 166.8,
185.2, 200.1\,.  {\bf 3C277.3} at 14.3$\mu$m ( LW3): $\pm$3.2,
$\pm$6.5, 11.6, 62.0, 93.0\,.  {\bf 3C288.1} at 12.0$\mu$m (LW10):
$\pm$18.5, $\pm$36.9, 62.6, 66.5, 93.9\,.  {\bf 3C293} at 12.0$\mu$m
(LW10): $\pm$47.4, $\pm$94.9, 170.8, 184.3, 276.4\,.  {\bf 3C295} at
6.7$\mu$m (LW2): $\pm$2.5, $\pm$5.0, 9.1, 9.6, 13.6\,.  {\bf 3C296} at
4.5$\mu$m (LW1): $\pm$37.8, $\pm$40.0, 79.9, 119.9, 143.9\,.  }
\end{figure*}

\clearpage
\setcounter{figure}{0}
\begin{figure*}
 {\epsfig{file=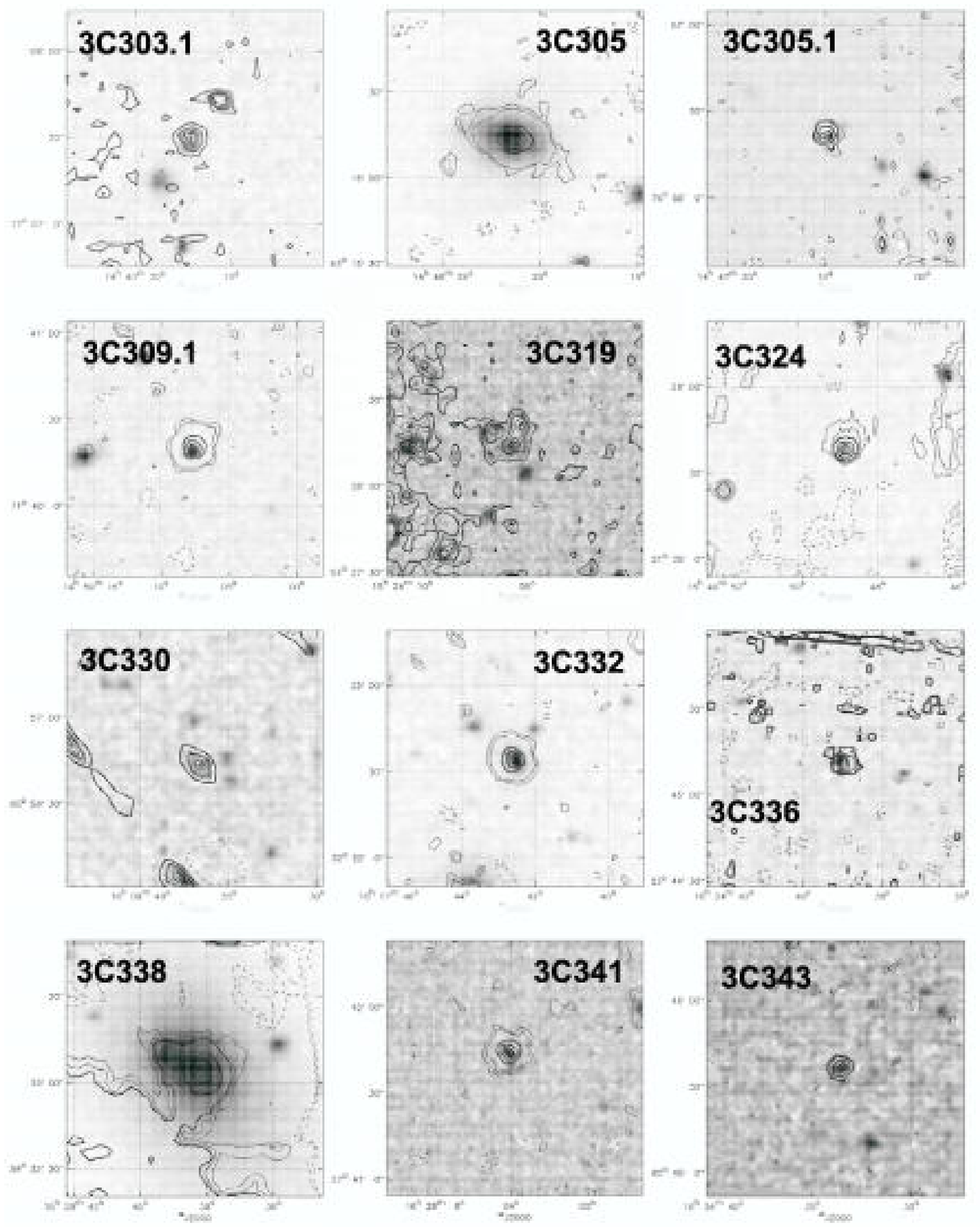}} \hfill
\caption{ -- continued --: {\bf 3C303.1} at 12.0$\mu$m (LW10):
$\pm$5.0, $\pm$10.0, 17.9, 117.5, 176.3\, {\bf 3C305.0} at 12.0$\mu$m
(LW10): $\pm$50.5, $\pm$101.0, 151.5, 181.8, 185.0\,.  {\bf 3C305.1}
at 12.0$\mu$m (LW10): $\pm$7.3, $\pm$14.7, 26.4, 73.4, 110.2\,.  {\bf
3C309.1} at 12.0$\mu$m (LW10): $\pm$39.8, $\pm$79.3, 118.9, 119.4,
143.3\,.  {\bf 3C319} at 12.0$\mu$m (LW10): $\pm$2.5, $\pm$4.9, 8.9,
109.5, 164.2\,.  {\bf 3C324} at 12.0$\mu$m (LW10): $\pm$7.8,
$\pm$15.7, 28.3, 355.4, 533.1\,.  {\bf 3C330} at 14.3$\mu$m (LW3):
$\pm$2.1, $\pm$4.3, 6.4, 7.7, 8.5\,.  {\bf 3C332} at 12.0$\mu$m
(LW10): $\pm$45.0, $\pm$89.9, 134.9, 161.8, 175.2\,.  {\bf 3C336} at
12.0$\mu$m (LW10): $\pm$9.7, $\pm$19.4, 34.9, 74.3, 111.4\,.  {\bf
3C338} at 6.7$\mu$m (LW2): $\pm$4.3, $\pm$7.9, 11.9, 13.0, 15.6\,.
{\bf 3C341} at 12.0$\mu$m (LW10): $\pm$14.4, $\pm$21.7, 73.4, 110.1,
132.2\,.  {\bf 3C343} at 12.0$\mu$m (LW10): $\pm$6.9, $\pm$13.8, 24.8,
101.4, 152.1\,.  }
\end{figure*}

\clearpage
\setcounter{figure}{0}
\begin{figure*}
 {\epsfig{file=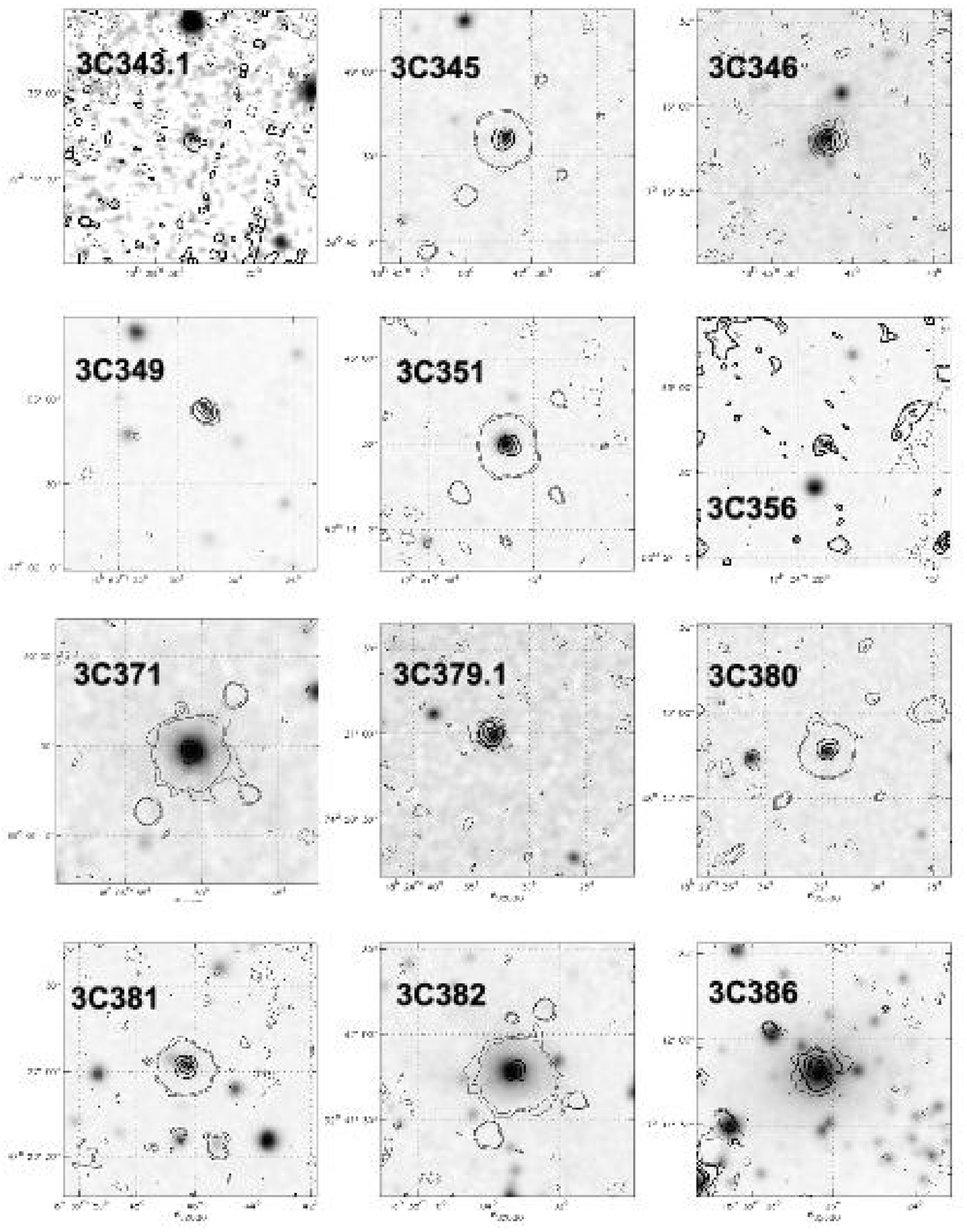}} \hfill
\caption{ -- continued --: {\bf 3C343.1} at 12.0$\mu$m (LW10):
$\pm$2.9, $\pm$5.7, 10.3, 58.9, 88.3\,.  {\bf 3C345} at 12.0$\mu$m
(LW10): $\pm$118.1, $\pm$177.1, 362.8, 544.2, 653.0 \,.  {\bf 3C346}
at 12.0$\mu$m (LW10): $\pm$20.2, $\pm$40.5, 72.9, 132.2, 198.3\,.
{\bf 3C349} at 12.0$\mu$m (LW10): $\pm$15.8, $\pm$31.6, 56.8, 106.2,
159.3\,.  {\bf 3C351} at 12.0$\mu$m (LW10): $\pm$13.2, $\pm$19.8,
349.5, 524.3, 629.2 \,.  {\bf 3C356} at 12.0$\mu$m (LW10): $\pm$4.6,
$\pm$9.2, 16.6, 113.8, 170.7\,.  {\bf 3C371} at 12.0$\mu$m (LW10):
$\pm$118.8, $\pm$178.2, 757.4, 1136.1, 1363.3\,.  {\bf 3C379.1} at
12.0$\mu$m (LW10): $\pm$9.7, $\pm$14.5, 31.3, 47.0, 56.3\,.  {\bf
3C380.0} at 12.0$\mu$m (LW10): $\pm$78.4, $\pm$101.1, 156.8, 235.2,
282.3\,.  {\bf 3C381} at 12.0$\mu$m (LW10): $\pm$91.0, $\pm$132.1,
198.2, 272.9, 327.4\,.  {\bf 3C382} at 12.0$\mu$m (LW10) : $\pm$138.8,
$\pm$208.1, 992.6, 1488.9, 1786.7\,.  {\bf 3C386} at 12.0 $\mu$m
(LW10): $\pm$8.8, $\pm$17.6, 31.7, 144.5, 216.7\,.  }
\end{figure*}

\clearpage

\setcounter{figure}{0}
\begin{figure*}
{\epsfig{file=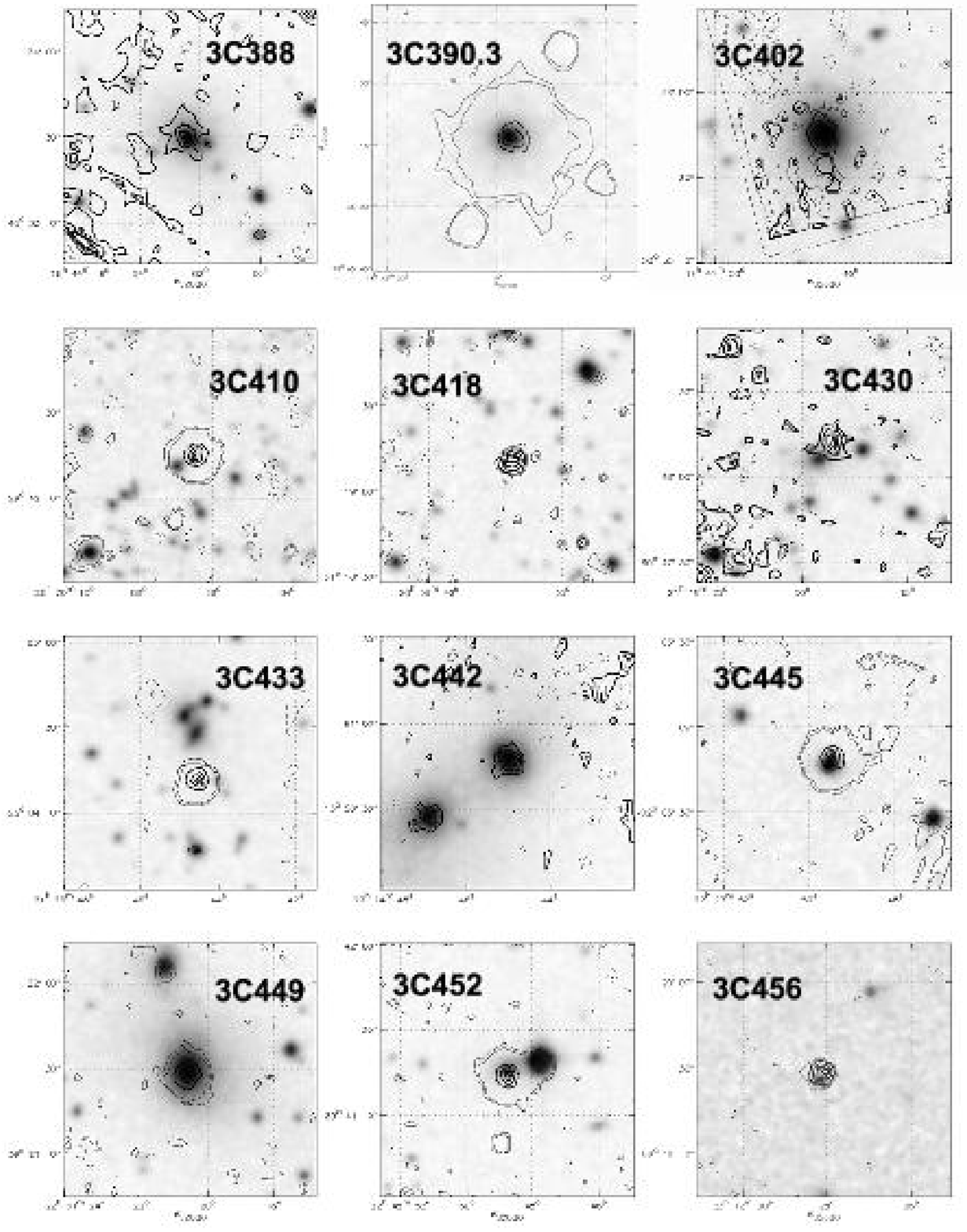}} \hfill
\caption{ -- continued --: {\bf 3C388} at 12.0$\mu$m (LW10): $\pm$3.4,
$\pm$6.8, 12.3, 110.0, 165.0\,.  {\bf 3C390.3} at 12.0 $\mu$m (LW10):
$\pm$119.1, $\pm$178.7, 1031.4, 1547.1, 1856.5\,.  {\bf 3C402} at
12.0$\mu$m (LW10): $\pm$12.1, $\pm$24.2, 43.6, 173.5, 260.3\,.  {\bf
3C410} at 12.0$\mu$m (LW10): $\pm$114.9, $\pm$143.2, 229.8, 344.6,
413.6\,.  {\bf 3C418} at 12.0$\mu$m (LW10): $\pm$16.0, $\pm$32.0,
57.6, 174.8, 262.2\,.  {\bf 3C430} at 12.0$\mu$m (LW10): $\pm$ 30,
$\pm$ 50, 60, 100, 150, 180\,.  {\bf 3C433} at 12.0$\mu$m (LW10):
$\pm$230.4, $\pm$232.0, 463.9, 695.9, 835.1\,.  {\bf 3C442, NGC7236
and NGC7237} at 12.0$\mu$m (LW10): $\pm$8.2, $\pm$16.4, 29.4, 178.2,
267.3\,.  {\bf 3C445} at 14.3$\mu$m (LW3): $\pm$258.5, $\pm$387.7,
1678.3, 2517.4, 3020.9\,.  {\bf 3C449} at 12.0$\mu$m (LW10):
$\pm$32.3, $\pm$64.7, 116.4, 164.6, 246.9\,.  {\bf 3C452} at
12.0$\mu$m (LW10): $\pm$101.4, $\pm$202.8, 304.2, 314.6, 365.1\,.
{\bf 3C456} at 12.0$\mu$m (LW10): $\pm$20.4, $\pm$30.7, 73.7, 110.5,
132.6\,.}
\end{figure*}

\clearpage

\setcounter{figure}{0}
\begin{figure*}
 {\epsfig{file=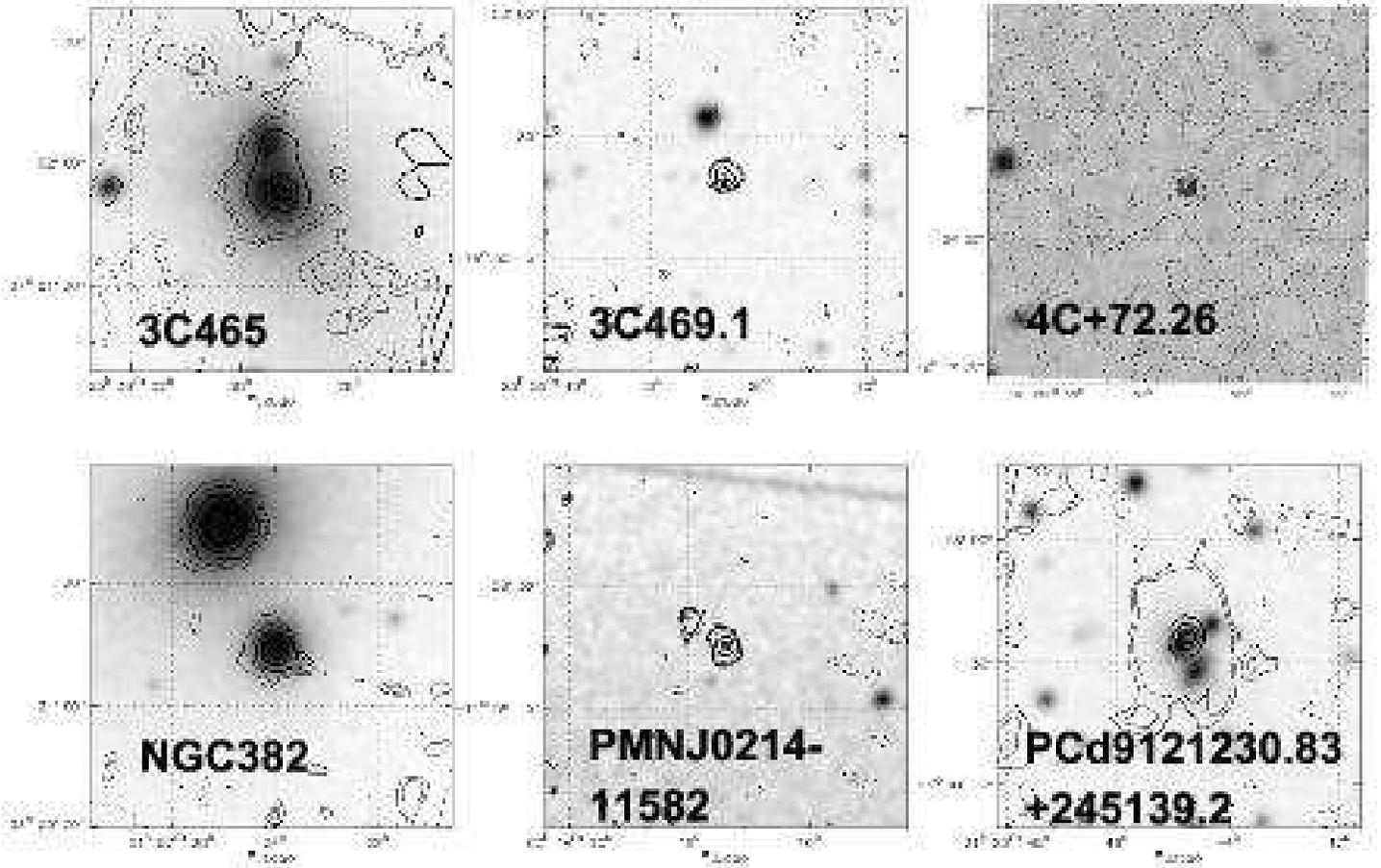}} \hfill
\caption{ -- continued --: {\bf 3C465} at 6.7$\mu$m (LW2): $\pm$9.2,
$\pm$9.9, 19.7, 29.6, 35.5\,.  {\bf 3C469.1} at 12.0$\mu$m (LW10):
$\pm$10.3, $\pm$20.7, 37.2, 181.7, 272.5\,.  {\bf 4C+72.26} at 12.0
$\mu$m (LW10): $\pm$0.4, $\pm$0.7, 1.3, 114.0, 171.1\,.  {\bf NGC382}
at 12.0$\mu$m (LW10): $\pm$27.3, $\pm$54.6, 98.3, 173.4, 260.1\,.
{\bf PMNJ0214$-$11582} at 12.0$\mu$m (LW10): $\pm$4.6, $\pm$9.1, 16.4,
250.0, 375.0\,.  {\bf PCd91212130.83+245139.2} at 6.7$\mu$m (LW2):
$\pm$48.2, $\pm$72.3, 235.9, 353.9, 424.6\,.  }
\end{figure*}

\begin{table*}
\caption{ISOCAM observations of 3CR sources ordered by TDT number.}
\label{tab.obs}
\begin{center}
\begin{tabular}{cccclrrrrrrr}
\hline
\hline
         & & & &  &  & & & &  &  & \\
        1&2&3&4& 5& 6&7&8&9&10&11&12 \\
         & & & &  &  & & & &  &  & \\
TDT & RA & Dec & Date & Filter & Lens & $M$ & $N$ & $dM$ & $dN$ & $N_{\rm {exp}}$ & $t_{\rm {int}}$ \\
 & J2000.0 & & UT & &  $''$ &  & &$''$ &$''$ & & $s$ \\
         & & & &  &  & & & &  &  & \\
\hline
    08800387 &  15 11 03.6  & +07 49 52.9  & 13.02.96 & LW10 & 6&   2 &    2 & 180& 180&    8 &  2 \\
    10200421 &  18 05 06.7  & +11 01 31.2  & 27.02.96 &  LW3 & 1.5 &   2 &    2 &  13.5 &  13&   51 &  2 \\
    & & & &   LW2 & 1.5 &   2 &    2 &  13.5 &  13&   51 &  2 \\
    10601408 &  16 28 38.6  & +39 33 07.0  & 02.03.96 &  LW3 & 3&   3 &    3 &   9&   9&   16 &  2 \\
   & &  &    &  LW1 & 3&   3 &    3 &   9&   9&   12 &  5\\
    12300106 &  09 55 51.7  & +69 40 45.9  & 19.03.96 & CVF & 3& 1 & 1 & 0 & 0 & 25 & 2 \\
    12300205 &  14 49 21.5  & +63 16 13.6  & 19.03.96 &  LW3 & 3&   3 &    3 &  15&  15&   49 &  2 \\
   & & &     &  LW2 & 1.5 &   3 &    3 &   7.5 &   7&   18 &  2 \\
    14301909 &  14 36 37.1  & +63 19 13.8  & 08.04.96 & LW10 & 3&   5 &    5 & 144& 144&   44 &  5\\
    14302010 &  14 36 37.1  & +63 19 13.8  & 08.04.96 & LW10 & 3&   5 &    5 & 144& 144&   44 &  5\\
    18001405 &  14 11 20.6  & +52 12 08.6  & 15.05.96 &  LW3 & 3&   4 &    4 & 144& 144&   30 & 10 \\
    & & &      &  LW2 & 3&   4 &    4 & 144& 144&   30 & 10 \\
    21305001 & 11 04 13.8  & +76 58 57.3  & 17.06.96 & CVF & 6& 1 & 1 & 0 & 0 & 14 & 10\\
    22201802 &  11 45 28.6  & +31 33 49.1  & 26.06.96 &  LW2 & 3&   5 &    5 &   9&   9&   13 & 10 \\
22400303 &  11 45 28.6  & +31 33 49.1  & 28.06.96 & LW10 & 3&   7 &   7 &   6&   6&   15 & 10 \\
22400201 &  11 45 28.6  & +31 33 49.1  & 27.06.96 &  LW1 & 3&   5 &   5 &   9&   9&   19 & 10 \\
    22801205 &  12 19 23.2  & +05 49 29.4  & 02.07.96 &  LW3 & 3&   3 &    3 &   9&   9&   12 &  2 \\
    & & &      &  LW2 & 3&   3 &    3 &   9&   9&   12 &  2 \\
    23100414 &  12 25 03.7  & +12 53 14.0  & 05.07.96 &  LW7 & 3&   5 &    5 &   9&   9&    4 &  5\\
    & & &     &  LW3 & 3&   5 &    5 &   9&   9&    4 &  5\\
    & & &     &  LW2 & 3&   5 &    5 &   9&   9&    3 & 20 \\
    23502406 &  12 25 03.6  & +12 53 14.1  & 09.07.96 &  LW1 & 6&   2 &    2 & 108& 108&   25 &  5\\
    23800308 &  12 30 49.4  & +12 23 28.1  & 12.07.96 &  LW1 & 6&   2 &    2 & 108& 108&   25 &  5\\
    & & &     &  LW2 & 3&   2 &    2 &  54&  54&   18 &  5\\
    & & &     &  LW3 & 6&   2 &    2 & 108& 108&   23 &  2 \\
    23901834 &  12 30 48.6  & +12 23 33.0  & 13.07.96 & LW10 & 1.5 &   3 &    3 &   4.5 &   4&   10 &  5\\
    24100504 &  12 28 46.8  &+02 08 07.9  & 14.07.96 & LW3 & 1.5 &   1 &    1 & 0 & 0 &   508 &  5\\
    & & &     & LW2 & 1.5 &   1 &    1 & 0 & 0 &   629 &  5\\
    24201104 &  12 45 38.4  & +03 23 20.3  & 16.07.96 & LW10 & 3&   5 &    5 & 144& 144&   27 &  5\\
24407233 &  13 42 13.2  & +60 21 42.0  & 18.07.96 & LW10 & 3&   2 &   2 &  60&  60&    5 & 10 \\
    24500106 &  12 54 12.0  & +27 37 33.5  & 18.07.96 &  LW3 & 3&   3 &    3 &  15&  15&   45 &  5\\
    & & &    &  LW2 & 1.5 &   3 &    3 &   7.5 &   7&   22 &  5\\
    27000606 &  14 16 52.8  & +10 48 23.0  & 13.08.96 &  LW2 & 3&   3 &    3 &   9&   9&   12 &  2 \\
    & & &    &  LW3 & 3&   3 &    3 &   9&   9&   12 &  2 \\
    27503920 &  14 36 43.1  & +15 44 13.0  & 17.08.96 & LW10 & 3&   5 &    5 & 144& 144&   32 &  5\\
    30300511 &  15 49 48.9  & +21 25 38.8  & 14.09.96 &  LW1 & 3&   5 &    5 &   9&   9&   36 & 10 \\
    30300612 &  15 49 48.9  & +21 25 38.8  & 14.09.96 &  LW2 & 3&   5 &    5 &   9&   9&   22 & 10 \\
30400442 &  16 24 39.0  & +23 45 13.0  & 15.09.96 & LW10 & 3&   2 &   2 &  60&  60&    5 & 10 \\
30492200 &  16 24 39.5  & +23 45 00.6  & 16.09.96 & LW2 & 6&   1 &   1 &  0&  0&    6 & 25.2 \\
    33600323 &  16 09 36.5  & +65 56 49.0  & 17.10.96 &  LW3 & 6&   4 &    4 & 504& 504&   24 & 10 \\
    & & &     &  LW2 & 6&   4 &    4 & 504& 504&   24 & 10 \\
    35200533 &  19 59 25.8  & +40 44 11.0  & 02.11.96 &  LW2 & 1.5 &   3 &    3 &   4.5 &   4&    9 & 10 \\
    35701635 &  21 23 44.5  & +25 04 10.0  & 08.11.96 &  LW2 & 3&   4 &    4 &   9&   9&    8 & 10 \\
    36701710 &  22 31 20.5  & +39 21 30.0  & 18.11.96 &  LW3 & 3&   3 &    3 &   9&   9&   13 &  2 \\
    & & &    &  LW2 & 3&   3 &    3 &   9&   9&   13 &  2 \\
   37403041 &  22 31 20.5  & +39 21 30.0  & 25.11.96 & LW10 & 6&   4 &   4 &  36&  36&    9 & 10 \\
37500303 &  23 16 34.5  & +04 05 17.4  & 25.11.96 & LW1 & 6&   1 &   1 &  0&  0&    12 & 5 \\
         &              &              &          & LW4 & 6&   1 &   1 &  0&  0&    12 & 5 \\
         &              &              &          & LW6 & 6&   1 &   1 &  0&  0&    12 & 5 \\
         &              &              &          & LW3 & 6&   1 &   1 &  0&  0&    12 & 5 \\
    38600922 &  16 24 39.5  & +23 45 00.6  & 10.07.97 & LW2 & 6&   1 &    1 &  0&  0&   7 &  25.2\\
    38800808 & 13 31 08.2  & +30 30 33.0  & 08.12.96 & LW10 & 3& 1 & 1 & 0 & 0 & 206 & 2 \\
    39501112 &  23 38 29.4  & +27 01 54.0  & 15.12.96 &  LW3 & 3&   3 &    3 &   9&   9&   16 &  2 \\
    & & &    &  LW2 & 3&   3 &    3 &   9&   9&   16 &  2 \\
         & & & &  &  & & & &  &  & \\
\hline
\end{tabular}
\end{center}
\end{table*}

%\clearpage

\setcounter{table}{0}
\begin{table*}
\caption{continued}
\begin{center}
\begin{tabular}{cccclrrrrrrr}
\hline
\hline
         & & & &  &  & & & &  &  & \\
        1&2&3&4& 5& 6&7&8&9&10&11&12 \\
         & & & &  &  & & & &  &  & \\
TDT & RA & Dec & Date & Filter & Lens & $M$ & $N$ & $dM$ & $dN$ & $N_{\rm {exp}}$ & $t_{\rm {int}}$ \\
 & J2000.0 & & UT & &  $''$ &  & &$''$ &$''$ & & $s$ \\
         & & & &  &  & & & &  &  & \\
\hline
    39901248 &  16 38 28.1  & +62 34 44.0  & 19.12.96 & LW10 & 3&   2 &    2 &  60&  60&    5 & 10 \\
    40100319 &  14 36 43.1  & +15 44 13.0  & 21.12.96 & LW10 & 3&   5 &    5 & 144& 144&   32 &  5\\
    40201422 &  01 07 27.4  & +32 24 08.0  & 22.12.96 &  LW2 & 6&   3 &    3 &  36&  36&    9 & 10 \\
    41907203 &  00 06 22.6  & $-$00 04 25.0  & 08.01.97 & LW10 & 3&   2 &    2 &  60&  60&    5 & 10 \\
    43502724 &  02 23 11.5  & +42 59 30.0  & 24.01.97 &  LW2 & 3&   2 &    3 &   9&   9&    8 & 10 \\
    43901804 &  01 37 41.2  & +33 09 35.3  & 29.01.97 &  LW3 & 1.5 &   2 &    2 &  22.5 &  22&   20 &  5\\
    45304902 &  02 23 11.5  & +42 59 30.0  & 11.02.97 &  LW3 & 3&   3 &    3 &   9&   9&   10 &  2 \\
    & & &   &  LW1 & 3&   3 &    3 &   9&   9&    8 &  5\\
46300408 &  14 49 21.5  & +63 16 14.0  & 21.02.97 &  LW1 & 3&   3 &   3 &   9&   9&    8 &  5\\
    46601603 &  03 03 15.0  & +16 26 16.0  & 24.02.97 &  LW2 & 3&   3 &    3 &   9&   9&    8 &  2 \\
    & & &     &  LW3 & 3&   3 &    3 &   9&   9&    9 &  2 \\
    & & &    &  LW1 & 3&   3 &    3 &   9&   9&    7 &  5\\
    47100710 &  18 38 25.4  & +17 11 52.0  & 01.03.97 &  LW2 & 3&   3 &    3 &   9&   9&   15 &  2 \\
    & & &   &  LW3 & 3&   3 &    3 &   9&   9&   15 &  2 \\
    49604639 &  19 41 42.8  & +50 36 36.7  & 26.03.97 & LW10 & 3&   3 &    3 &  72&  72&   15 &  5\\
    50304334 &  18 33 45.9  & +47 27 01.2  & 02.04.97 & LW10 & 3&   3 &    3 &  72&  72&   15 &  5\\
    51100561 &  16 28 38.3  & +39 33 04.5  & 10.04.97 & LW10 & 3&   3 &    3 &  72&  72&   15 &  5\\
    51100679 &  16 42 58.7  & +39 48 36.7  & 10.04.97 & LW10 & 3&   3 &    3 &  72&  72&   15 &  5\\
    51301035 &  18 38 26.3  & +17 11 49.5  & 12.04.97 & LW10 & 3&   3 &    3 &  72&  72&   15 &  5\\
    51400760 &  14 49 21.8  & +63 16 14.0  & 13.04.97 & LW10 & 3&   3 &    3 &  72&  72&   15 &  5\\
    51800103 &  17 14 11.9  & +50 16 60.0  & 17.04.97 & LW10 & 6&  11 &   11 & 198& 198&   12 &  5\\
    51800667 &  17 10 44.0  & +46 01 29.1  & 17.04.97 & LW10 & 3&   3 &    3 &  72&  72&   15 &  5\\
    51800831 &  17 28 20.1  & +31 46 01.9  & 17.04.97 & LW10 & 3&   3 &    3 &  72&  72&   15 &  5\\
    52100668 &  17 24 19.3  & +50 57 36.1  & 20.04.97 & LW10 & 3&   3 &    3 &  72&  72&   15 &  5\\
    52100804 &  17 14 11.9  & +50 16 60.0  & 20.04.97 & LW10 & 6&  11 &   11 & 198& 198&   12 &  5\\
    52201845 &  21 23 44.0  & +25 04 31.1  & 21.04.97 & LW10 & 3&   3 &    3 &  72&  72&   15 &  5\\
    52201948 &  21 44 11.6  & +28 10 18.7  & 21.04.97 & LW10 & 3&   3 &    3 &  72&  72&   15 &  5\\
    52500130 &  17 04 41.3  & +60 44 30.1  & 24.04.97 & LW10 & 3&   3 &    3 &  72&  72&   15 &  5\\
    52500429 &  16 59 28.8  & +47 02 55.4  & 24.04.97 & LW10 & 3&   3 &    3 &  72&  72&   15 &  5\\
    52500562 &  18 35 03.3  & +32 41 46.9  & 24.04.97 & LW10 & 3&   3 &    3 &  72&  72&   15 &  5\\
    52900115 &  14 43 14.6  & +77 07 27.6  & 28.04.97 & LW10 & 3&   3 &    3 &  72&  72&   15 &  5\\
    52900417 &  15 10 23.0  & +70 45 52.8  & 28.04.97 & LW10 & 3&   3 &    3 &  72&  72&   15 &  5\\
    52901232 &  18 06 50.5  & +69 49 29.9  & 28.04.97 & LW10 & 3&   3 &    3 &  72&  72&   15 &  5\\
    52901433 &  18 24 32.8  & +74 20 57.2  & 28.04.97 & LW10 & 3&   3 &    3 &  72&  72&   15 &  5\\
    52901537 &  18 42 08.7  & +79 46 17.8  & 28.04.97 & LW10 & 3&   3 &    3 &  72&  72&   15 &  5\\
    52902663 &  21 04 06.3  & +76 33 11.8  & 28.04.97 & LW10 & 3&   3 &    3 &  72&  72&   15 &  5\\
    53200838 &  19 40 24.9  & +60 41 36.7  & 01.05.97 & LW10 & 3&   3 &    3 &  72&  72&   15 &  5\\
    53503041 &  20 20 06.5  & +29 42 13.6  & 04.05.97 & LW10 & 3&   3 &    3 &  72&  72&   15 &  5\\
    53503146 &  21 23 16.2  & +15 48 05.6  & 04.05.97 & LW10 & 3&   3 &    3 &  72&  72&   15 &  5\\
    53503288 &  21 47 25.1  & +15 20 33.8  & 04.05.97 & LW10 & 3&   3 &    3 &  72&  72&   15 &  5\\
    53702750 &  22 14 46.8  & +13 50 27.1  & 06.05.97 & LW10 & 3&   3 &    3 &  72&  72&   15 &  5\\
    53702911 &  22 14 46.9  & +13 50 27.0  & 06.05.97 &  LW3 & 3&   3 &    3 &   9&   9&   15 &  2 \\
    53702911 &  22 14 46.9  & +13 50 27.0  & 06.05.97 &  LW2 & 3&   3 &    3 &   9&   9&   15 &  2 \\
    54000949 &  21 55 52.1  & +38 00 29.4  & 09.05.97 & LW10 & 3&   3 &    3 &  72&  72&   15 &  5\\
    54100619 &  15 24 04.8  & +54 28 05.2  & 10.05.97 & LW10 & 3&   3 &    3 &  72&  72&   15 &  5\\
    54100836 &  18 44 02.3  & +45 33 30.3  & 10.05.97 & LW10 & 3&   3 &    3 &  72&  72&   15 &  5\\
    54100969 &  18 29 31.7  & +48 44 46.3  & 10.05.97 & LW10 & 3&   3 &    3 &  72&  72&   15 &  5\\
    54101870 &  20 38 36.8  & +51 19 12.1  & 10.05.97 & LW10 & 3&   3 &    3 &  72&  72&   15 &  5\\
    54301344 &  21 18 19.0  & +60 48 08.2  & 12.05.97 & LW10 & 3&   3 &    3 &  72&  72&   15 &  5\\
    54500518 &  22 23 49.5  & $-$02 06 12.3  & 14.05.97 &  LW2 & 3&   1 &    1 &   0 &   0 &   43 &  2 \\
    & & &     &  LW3 & 3&   1 &    1 &   0 &   0 &   49 &  2 \\
    54700854 &  23 12 28.1  & +09 19 27.6  & 16.05.97 & LW10 & 3&   3 &    3 &  72&  72&   15 &  5\\
    54801053 &  22 45 48.7  & +39 41 16.6  & 17.05.97 & LW10 & 3&   3 &    3 &  72&  72&   15 &  5\\
54801112 &  22 31 20.5  & +39 21 30.0  & 17.05.97 &  LW1 & 3&   3 &   3 &   9&   9&    8 &  5\\
         & & & &  &  & & & &  &  & \\
\hline
\end{tabular}
\end{center}
\end{table*}

%\clearpage

\setcounter{table}{0}
\begin{table*}
\caption{continued}
\begin{center}
\begin{tabular}{cccclrrrrrrr}
\hline
\hline
         & & & &  &  & & & &  &  & \\
        1&2&3&4& 5& 6&7&8&9&10&11&12 \\
         & & & &  &  & & & &  &  & \\
TDT & RA & Dec & Date & Filter & Lens & $M$ & $N$ & $dM$ & $dN$ & $N_{\rm {exp}}$ & $t_{\rm {int}}$ \\
 & J2000.0 & & UT & &  $''$ &  & &$''$ &$''$ & & $s$ \\
         & & & &  &  & & & &  &  & \\
\hline
    54801252 &  22 31 20.5  & +39 21 29.4  & 17.05.97 & LW10 & 3&   3 &    3 &  72&  72&   15 &  5\\
    55500476 &  18 06 50.8  & +69 49 30.0  & 24.05.97 &  LW3 & 3&   1 &    1 &   0 &   0 &  109 &  2 \\
    56001074 &  22 50 32.9  & +71 29 19.0  & 29.05.97 & LW10 & 3&   3 &    3 &  72&  72&   15 &  5\\
    56201064 &  23 55 23.0  & +79 55 18.7  & 31.05.97 & LW10 & 3&   3 &    3 &  72&  72&   15 &  5\\
    56201411 &  02 22 36.2  & +86 19 07.6  & 31.05.97 & LW10 & 3&   3 &    3 &  72&  72&   15 &  5\\
    56500458 &  23 38 29.4  & +27 01 53.3  & 03.06.97 & LW10 & 3&   3 &    3 &  72&  72&   15 &  5\\
    56501357 &  23 21 28.5  & +23 46 47.2  & 03.06.97 & LW10 & 3&   3 &    3 &  72&  72&   15 &  5\\
    57502102 &  00 38 20.5  & $-$02 07 40.8  & 13.06.97 & LW10 & 3&   3 &    3 &  72&  72&   15 &  5\\
    58703793 &  01 07 24.9  & +32 24 44.9  & 25.06.97 & LW10 & 3&   3 &    3 &  72&  72&   15 &  5\\
    58703801 &  01 07 25.0  & +32 24 45.2  & 25.06.97 &  LW3 & 3&   3 &    3 &   9&   9&    9 &  2 \\
    59501620 &  15 31 25.4  & +35 33 40.5  & 03.07.97 & LW10 & 3&   3 &    3 &  72&  72&   15 &  5\\
    59601514 &  02 14 17.4  & $-$11 58 46.2  & 04.07.97 & LW10 & 3&   8 &    8 &  36&  36&   13 &  5\\
    59601615 &  02 14 17.4  & $-$11 58 46.2  & 04.07.97 & LW10 & 3&   8 &    8 &  36&  36&   12 &  5\\
    59702305 &  00 43 09.2  & +52 03 33.7  & 05.07.97 & LW10 & 3&   3 &    3 &  72&  72&   15 &  5\\
    59702607 &  01 09 44.3  & +73 11 57.6  & 05.07.97 & LW10 & 3&   3 &    3 &  72&  72&   15 &  5\\
    60001373 &  16 09 34.6  & +65 56 37.7  & 08.07.97 & LW10 & 3&   3 &    3 &  72&  72&   15 &  5\\
    60001472 &  14 59 07.5  & +71 40 19.5  & 08.07.97 & LW10 & 3&   3 &    3 &  72&  72&   15 &  5\\
    60009220 &  14 59 13.5  & +71 40 26.9  & 08.07.97 & LW2 & 6&   1 &    1 &  0&  0&   7 &  25.2\\
    60201626 &  16 28 03.8  & +27 41 38.6  & 10.07.97 & LW10 & 3&   3 &    3 &  72&  72&   15 &  5\\
    60201725 &  16 17 42.8  & +32 22 28.6  & 10.07.97 & LW10 & 3&   3 &    3 &  72&  72&   15 &  5\\
    61503617 &  03 19 48.2  & +41 30 42.4  & 23.07.97 &  LW2 & 6&   2 &    2 & 108& 108&   25 &  5\\
    61503807 &  03 10 00.3  & +17 05 54.0  & 23.07.97 & LW10 & 3&   1 &    1 &   0 &   0 &   61 &  2 \\
    61701307 &  13 52 17.8  & +31 26 47.0  & 25.07.97 &  LW2 & 3&   3 &    3 &   9&   9&    8 &  2 \\
& & &    &  LW3 & 3&   3 &    3 &   9&   9&    9 &  2 \\
    61701413 &  13 52 17.8  & +31 26 46.4  & 25.07.97 & LW10 & 3&   3 &    3 &  72&  72&   15 &  5\\
    61900565 &  16 34 33.7  & +62 45 36.2  & 27.07.97 & LW10 & 3&   3 &    3 &  72&  72&   15 &  5\\
    61901003 &  00 40 50.3  & +10 03 23.4  & 27.07.97 & LW10 & 3&   3 &    3 &  72&  72&   15 &  5\\
63301602 & 02 42 40.7  & $-$00 00 47.3  & 10.08.97 & LW1 & 3& 2 & 2 & 36 & 36 & 247 & 2 \\
63301902 & 02 42 40.7  & $-$00 00 47.3  & 10.08.97 & LW5 & 3& 2 & 2 & 36 & 36 & 247 & 2 \\
63302202 & 02 42 40.7  & $-$00 00 47.3  & 10.08.97 & LW9 & 3& 2 & 2 & 36 & 36 & 266 & 2 \\
    63302405 &  03 58 54.7  & +10 25 47.0  & 10.08.97 &  LW2 & 3&   3 &    3 &   9&   9&    9 &  2 \\
    & & &    &  LW3 & 3&   3 &    3 &   9&   9&    8 &  2 \\
    64001327 &  16 43 48.7  & +17 15 49.3  & 17.08.97 & LW10 & 3&   3 &    3 &  72&  72&   15 &  5\\
65800208 & 15 31 42.7  & +24 04 24.7  & 03.09.97 & CVF & 3& 1 & 1 & 0 & 0 & 33 & 2 \\
    65901032 &  03 18 15.7  & +41 51 27.0  & 05.09.97 &  LW2 & 3&   4 &    4 &   9&   9&    9 & 10 \\
    65901304 &  03 18 15.7  & +41 51 27.0  & 05.09.97 &  LW3 & 3&   3 &    3 &   9&   9&    9 &  2 \\
    70101081 &  00 16 31.1  & +79 16 41.5  & 16.10.97 & LW10 & 3&   3 &    3 &  72&  72&   15 &  5\\
    71200283 &  18 05 06.3  & +11 01 31.2  & 27.10.97 &  LW2 & 1.5 &   1 &    1 &   0 &   0 &   83 &  5\\
    & & &     &  LW3 & 1.5 &   1 &    1 &   0 &   0 &   80 &  5\\
    71702771 &  14 47 08.9  & +76 56 19.6  & 01.11.97 & LW10 & 3&   3 &    3 &  72&  72&   15 &  5\\
    74101506 &  19 08 23.7  & +72 20 11.8  & 25.11.97 & LW10 & 3&   8 &    8 &  36&  36&   18 &  5\\
    75300580 &  16 09 36.5  & +65 56 42.7  & 07.12.97 &  LW3 & 6&  20 &    3 & 228&  18&   15 &  5\\
    75300681 &  16 09 36.5  & +65 56 42.7  & 07.12.97 &  LW3 & 6&  20 &    3 & 228&  18&   15 &  5\\
    75300779 &  16 09 36.5  & +65 56 42.7  & 07.12.97 &  LW3 & 6&  20 &    3 & 228&  18&   15 &  5\\
    75400182 &  16 09 36.5  & +65 56 42.7  & 08.12.97 &  LW3 & 6&  20 &    3 & 228&  18&   15 &  5\\
    78500882 &  00 50 56.2  & +51 12 03.8  & 08.01.98 & LW10 & 3&   3 &    3 &  72&  72&   15 &  5\\
    80201405 &  03 18 12.1  & $-$25 35 09.0  & 25.01.98 & LW10 & 3&   6 &    6 &  72&  72&   15 &  5\\
    80201507 &  03 18 12.1  & $-$25 35 09.0  & 25.01.98 & LW10 & 3&   6 &    6 &  72&  72&   14 &  5\\
    80801283 &  00 34 14.7  & +39 24 13.7  & 31.01.98 & LW10 & 3&   3 &    3 &  72&  72&   15 &  5\\
         & & & &  &  & & & &  &  & \\
\hline
\end{tabular}
\end{center}
\end{table*}


\begin{thebibliography}{}

\bibitem{} Andreani P., Fosbury R. A. E., van Bemmel I., Freudling,
W., 2002, A\&A 381, 389

\bibitem{} Akujor C.E., Ludke E., Browne I.W.A., Leahy J.P.,
Garrington S.T., et al., 1994, A\&A 105,247

\bibitem{} Akujor C.E.,  Garrington S. T., 1995, A\&AS 112, 235

%\bibitem{} Antonucci R.R.J., Miller J.S., 1985, ApJ 297621 

\bibitem{} Axon D.J., Capetti A., Fanti R., Morganti R., 2000, AJ 120, 2284

\bibitem{} Barthel P.D., 1999, ApJ 336, 606

\bibitem{} Baum S.A., O'Dea C.P., Giovannini G., Biretta J., Cotton,
W.B., et al. , 1997, ApJ 483, 178

\bibitem{} Best P.N., R\"ottgering H.J.A., Bremer M.N., et al., 1998,
MNRAS 301, L15

\bibitem{} Best P.N., Longair M.S., R\"ottgering H.J.A., 1998,
MNRAS 295, 549


%\bibitem{} Biviano A., Sauvage M., Gallais P. et al., 1998, ISOCAM
%Dark Current Calibration Report,
%{http://www.iso.vilspa.esa.es/users/expl\_lib/CAM\_list.html}.


\bibitem{} Blommaert J., Siebenmorgen R., Coulais A., et al., 2001,
``ISO Handbook Volume III (CAM)'', SAI-99-057/Dc,
{http://www.iso.vilspa.esa.es}


\bibitem{} Boisse P., Le Brun V., Bergeron J., Deharveng J.-M., 1998,
A\&A 333, 841

\bibitem{} Brunetti G., Setti G., Comastri A, 1997, A\&A 325, 898



\bibitem{} Canalizo G., Stockton A., 2000, ApJ 528, 201

\bibitem{} Capetti A., De Ruitter H.R., Fanti R., et al., 2000, A\&A 362, 871

\bibitem{} Cesarsky C., Abergel A., Agn\`ese P. et al., 1996, A\&A 315, L32

\bibitem{} Cimatti A., Freudling W., Rottgering H.J.A., Ivison R.J.,
Mazzei P., 1998, A\&A 329, 399

\bibitem{} Chambers K.C., Miley G.K., van Breugel W.J.M., Huang J.-S.,
Trentham A., 1996, ApJS 106, 247

%\bibitem{} Chidi E. A., Garrington S.T., 1995, A\&AS 112, 235

\bibitem{} Colbert J.W., Mulchaey J.S., Zabludoff A.I., 2001, AJ 121, 808

\bibitem{} Condon, W. D., Greisen E.W., Yin Q.F., Perley R.A., et al.,
1998, AJ 115, 1693

\bibitem{} Contini M., Contini T., 2003, MNRAS 342, 299

\bibitem{} Coulais A., Abergel A., 2000, A\&AAS 141, 533

\bibitem{} Crawford C.S., Fabian A.C., 1996, MNRAS 281, 5


%\bibitem{} de Breuck C., van Breugel W., Rottgering H., Stern D., Miley G., et al., 2001, AJ 121, 1241

\bibitem{} de Koff S., Baum S.A., Sparks W.B., Golombek D., Biretta
J., et al., 1996, ApJS 107, 621

\bibitem{} de Koff S., Best P., Baum S.A., Sparks W., Rottgering H.,
et al., 2000, ApJS 129, 33

\bibitem{} de Juan L., Colina L., Golombek D.,  1996, A\&A 305, 776

\bibitem{} de Vaucouleurs G., de Vaucouleurs A. , Corwin H.G., 1976,
Second reference catalogue of bright galaxies, Austin University of
Texas, Press.

\bibitem{} de Vries W.H., O'Dea C.P., Baum S.A., Sparks W.B., Biretta
J., 1997, ApJS 110, 191


\bibitem{} Djorgovski S., Spinrad H., McCarthy P., Dickinson M., van
Breugel W., et al., 1988, AJ 96, 836


\bibitem{} Draper P.W., Scarrott S.M., Tadhunter C.N., 1993, MNRAS 262, 1029

\bibitem{} Dwek E., Smith R.K., 1996, ApJ 459, 686


%\bibitem{} Efstathiou A., Hough J.H.,  Young S., 1995, MNRAS 277, 1134

\bibitem{} Efstathiou A., Rowan--Robinson, 1995, MNRAS 273, 649

\bibitem{} Eracleous M., Halpern J.P., 1994, ApJS 90, 1

%\bibitem{} Evans A.S., Frayer D.T., Surace J.A., Sanders D.B., 2001, AJ 121, 3285

\bibitem{} Fanti C., Pozzi F, Fanti R., et al., 2000, A\&A 358, 499

\bibitem{} Farrah D., Afonso J., Efstathiou, A., et al., 2003, MNRAS
343, 585

\bibitem{} Fernini I, Burns J.O., Bridle A.H., Perley R.A., 1993, AJ
105, 1690

\bibitem{} Ford H. C., Tsvetanov Z. I., Kriss G. A., Harms R., Dressel
L., 1994, A\&AS 184, 6402


\bibitem{} Freudling W., Siebenmorgen R., Haas M., 2003, ApJL 599, 13

\bibitem{} Gear W.K., Stevens J.A., Hughes D.H., et al. 1994, MNRAS
267, 167
 

\bibitem{} Giovannini G., Feretti L., Gregorini L., Parma P., 1998,
A\&A 199, 73


\bibitem{} Giovannini G., Cotton W.D., Feretti L., Lara L., Venturi
T., 2001, ApJ 552, 508


\bibitem{} Granato G.L., Danese L., 1994, MNRAS 268, 235

\bibitem{} Hardcastle M.J., Worrall D.M., 1999, MNRAS 309, 969

%\bibitem{} Hardcastle M. J., Alexander P., Pooley G.G., Riley J.M., 1999, MNRAS 317, 120


\bibitem{} Haas M., M\"uller S. A. H., Bertoldi F., et al., 2004, A\&A
submitted

\bibitem{} Harms R.,  Ford H.,  Tsvetanov Z. I., et al., 1994, ApJ 435, 35

\bibitem{} Harvanek M., Hardcastle M.J., 1998, ApJS 119, 25

\bibitem{} Heckman T.M., O'Dea C.P., Baum S.A., and Laurikainen
E. 1994, ApJ 428, 65

%\bibitem{} Hildebrand R., Whitcomb S., Winston R., et al., 1977, ApJ 216, 698

\bibitem{} Jackson N.,  Rawlings S. 1997, MNRAS 286, 241

\bibitem{} Jackson N., Beswick R.J., Pedlar A., et al., 2003, MNRAS 338, 643

\bibitem{} Jaffe W., Ford H., Ferrarese L., van den Bosch F.,
O'Connell R. W., 1996, ApJ 460, 214


\bibitem{} Katajainen S., Takalo L. O., Sillanp''a''a A., 2000 A\&AS, 143, 357

\bibitem{} Kotanyi C.G., Ekers R.D., 1979, A\&A 73, L1


%\bibitem{} Koyama K., Inoue H., Tanaka Y., Awaki H., Takano S., Ohashi  T., Matsuoka M., 1989, PASP 41, 731  

%\bibitem{} Krolik J.H., Kallmann T.R., 1987, ApJ 320, L5

\bibitem{} Kr\"ugel E., Chini R., Klein U., et al., 1990, A\&A 240, 232

\bibitem{} Kr\"ugel E., Siebenmorgen R., 1994, A\&A 288, 929 


%\bibitem{} Kuehr H., Witzel A., Pauliny-Toth I.I.K., Nauber U., 1981, A\&AS 45, 367


\bibitem{} Laing R.A., Jenkins C.R., Wall J.V., Unger S.W., 1994, The
First Stromlo Symposium: The Physics of Active Galaxies. ASP
Conference Series, Vol. 54, 1994, G.V., 201

\bibitem{} Leahy J.P., Perley R.A., 1991, AJ 102, 537

%\bibitem{} Le Brun V., Bergeron J., 1998, A\&A 332, 814

\bibitem{} Leech K., Pollock A.M.T. , 2001, ``ISO Handbok Vol. II'',
SAI-99-082/Dc, {http://www.iso.vilspa.esa.es}

\bibitem{} Lehnert M.D., Miley G.K., Sparks W.B., Baum S.A., Biretta
J., 1999, ApJS 123, 351

\bibitem{} Leon S., Lim J., Combes F., Van-Trung D., 2001, astroph/0107498


\bibitem{} Laor A., Draine B.T., 1993, ApJ 402, 441

\bibitem{} Ludke E., Garrington S.T., Spencer R.E., Akujor C.E.,
Muxlow T. W. B., 1998, MNRAS 299, 467

\bibitem{} Martel A.R., Braun. S.A., Sparks W.B., Wyckoff E., Biretta
J.A., et al., 1999, ApJS 122, 81

\bibitem{} McCarthy P. J., Spinrad H.,  van Breugel, W, 1995, ApJS 99, 27

\bibitem{} McCarthy P. J., Baum S.A., Spinrad H., 1996, ApJS 106, 281

\bibitem{} McCarthy P. J., Miley G.K., de Koff, et al., 1997, ApJS 112, 415

\bibitem{} Meisenheimer K., Haas M., M\"uller S. A. H., et al., 2001,
A\&A 372, 719

\bibitem{} Miller J.S., 1981, PASP 93, 681

\bibitem{} Morabito D.D., Preston R.A., Slade M.A., Jauncey D.L.,
1982, AJ 87, 1639


\bibitem{} Nenkova M., Ivezic Z., Elitzur M., 2002, ApJ 355, 456

\bibitem{} Neugebauer G., Green R. F., Matthews K., Schmidt M., Soifer
B. T., et al., 1987, ApJS 63, 615

\bibitem{} O'Dea P., de Vries W., Biretta J.A., Baum S.A., 1999, AJ 117, 1143

\bibitem{} Okumura K., ISOCAM PSF Report, 1998.
http://www.iso.vilspa.esa.es/users/expl\_lib/CAM\_list.html


\bibitem{} Ott S., Abergel A., Altieri B., et al., 1996, ASAP
Conference series, Vol. 125, 1997

\bibitem{} Ott S., Pollock A., Siebenmorgen R., 2000, ``The ISOCAM Parallel
Mode'', Proc. of a Ringberg Workshop, ISO Surveys of a Dusty
Universe, eds: D. Lemke, W. Stickel, K. Wilke, Lecture Notes in
Physics, Springer (ISBN 3-540-67479-9), p. 289 

\bibitem{} Peng B., Kraus A., Kirchbaum T.P., Witzel A., 2000, A\&AS, 145, 1

\bibitem{} Penston M.V., Cannon R.O., 1970, R. Obs. Bull. 159, 85 

\bibitem{} Pier A. P., Krolik J.H., 1993, ApJ 418, 673  

\bibitem{} Popescu C.C., Tuffs R.J., Kylafis N.D., Madore B.F., 2004,
A\&A 414, 45

%\bibitem{} Polletta M., Courvoisier T. J.--L., Hooper E. J., Wilkes B.J., 2000, A\&A 362, 75

%\bibitem{} Prieto A., 1996, MNRAS 282, 421

\bibitem{} Rantakyroe F.T., Baath L.B., Backer D.C., et al., 1998, A\&AS 131,

\bibitem{} Ridgway S.E., Stockton A., 1997, AJ 114, 511.

\bibitem{} Robson E.I., Stevens J.A., Jenness T., 2001, MNRAS 327, 751

\bibitem{} Roche N., Eales S.A., 1999, MNRAS 317, 120

\bibitem{} Roettgering H.J.A., Lacy M., Miley G. K., Chambers K. C.,
Saunders R.R, 1994, A\&AS 108, 79

\bibitem{} Roman P., Ott S., 1999, Report on the behaviour of ISOCAM
LW darks, ESA Technical Report,
{http://www.iso.vilspa.esa.es/users/expl\_lib/CAM\_list.html}.


\bibitem{} Rowan--Robinson M., 2000, MNRAS 316, 885

%\bibitem{} Savage B.D., Wakker B., Jannuzi B.T., Bahcall J.N., Bergeron J., et al., 2000, ApJS 129, 563

\bibitem{} Schmidt M., 1963, Nature 197, 1040


\bibitem{} Siebenmorgen R., Kr\"ugel E., Mathis J.S., 1992, A\&A 266, 501

\bibitem{} Siebenmorgen R., Moorwood A., Freudling W., K\"aufl H.U.,
1997, A\&A 325, 450. 

\bibitem{} Siebenmorgen R., Kr\"ugel E., Laureijs R., 2001, A\&A 377, 735

\bibitem{} Siebenmorgen R., Kr\"ugel E., Spoon H.W.W., 2004, A\&A 414, 123 


\bibitem{} Siebert J., Matsuoka M., Brinkmann W., Cappi M., Mihara T.,
et al., 1996, A\&A 307, 8

\bibitem{} Simpson C., Rawlings S., Lacy M., 1999, MNRAS 306, 828

%\bibitem{} Smith E.P., Heckman T. M., 1989, ApJ 341, 658

\bibitem{} Spinoglio L., Andreani P., Malkan M. A., 2002, ApJ 572, 105

\bibitem{} Spinrad H., Djorgovski S., Marr J., Aguilar L., 1985, PASP 97, 932

\bibitem{} Starck, J.L., Siebenmorgen, R. \& Gredel, R., 1997, ApJ 482,
1011-1020.

\bibitem{} Starck J.L, Abergel A., Aussel H., Sauvage M., Gastaud R., et
al., 1999, A\&AAS 134, 135-148.

\bibitem{} Steppe H.,  Salter C. J., Chini R., et al., 1988, A\&AS 75, 317

\bibitem{} Steppe H., Jeyakumar S., Saikia D. J., Salter C. J., 1995,
A\&AS 113, 409

\bibitem{} Sudou H.,Iguchi S., Murata Y., Taniguchi Y., 2003, Science
300, 1263

%\bibitem{} Tran D. 1998, PhD thesis, Paris: University of Paris XI 

\bibitem{} Worrall D. M., Birkinshaw M.,  2000, ApJ 530, 719 

\bibitem{} Tadhunter C.N, Morganti R., di Serego Alighieri S., Fosbury
R. A. E., Danziger I.J., 1993, MNRAS 263, 999

\bibitem{} Taylor G. L., Dunlop J. S., Hughes D. H., Robson E. I.,
1996, MNRAS 283, 930

\bibitem{} Urry, C. M.,  Padovani P.,  1995, PASP 107, 803

\bibitem{} Van Bemmel I.M., Barthel P.D., de Graauw T., 2001, A\&A 368, 414

%\bibitem{} Van Ojik R., Rottgering H.J.A., Miley G.K., Bremer M.N., Macchetto F., et al., 1994, A\&A 289, 54

%\bibitem{} Verdoes Kleijn G.A., Baum S.A., de Zeeuw P.T., O'Dea C.P., 2002, astroph/0112356v2

\bibitem{} Veron-Cetty M.-P., Veron, P., 2000, A Catalogue of quasars
and active nuclei, 9th edition, published by ESO,
{http://www.iso.vilspa.esa.es/users/expl\_lib/CAM\_list.html}.

%\bibitem{} Whitmore B.C., Sparks W.B., Lucas R.A., Macchetto F. D., Biretta J.A., 1995, ApJ 454, 73

\bibitem{} Xu C., Baum S.A., O'Dea C., Wrobel J.M., Condon J.J., 2000,
AJ 120, 2950

\bibitem{} Yee H.K.C., Oke J.B., 1978, ApJ 226, 753

\bibitem{} Zirbel E.L., Baum S. A., 1998, ApJS 114, 177

\end{thebibliography}
\end{document}